\documentclass[]{cit_lab_mfr}

\usepackage{verbatim}
\usepackage{wrapfig}
\usepackage{floatrow}
\usepackage{placeins}
\usepackage{algorithm}

\usepackage[page,header]{appendix}


\usepackage{silence}
\usepackage{adjustbox}
\usepackage{array}
\usepackage{tabularx}
\usepackage{pifont}
\usepackage{amsmath}
\usepackage{booktabs}
\usepackage{longtable}
\usepackage{multirow}
\usepackage{CJKutf8}

\newcolumntype{Y}{>{\centering\arraybackslash}X}

\definecolor{takeawaycolor}{RGB}{0,51,102}

\newtcolorbox{takeawaybox}[1][]{
  colback=takeawaycolor!5,
  colframe=takeawaycolor,
  fonttitle=\bfseries,
  title={#1},
  boxrule=0.8pt,
  arc=2pt,
  left=6pt,
  right=6pt,
  top=4pt,
  bottom=4pt,
  fontupper=\small\rmfamily
}

\definecolor{promptbg}{RGB}{0,51,102}
\definecolor{promptborder}{RGB}{0,51,102}
\definecolor{figurethreeblue}{RGB}{61,107,168}
\colorlet{tableaccent}{figurethreeblue!13}
\colorlet{metricaccent}{figurethreeblue!7}
\colorlet{tablegroup}{tableaccent}
\colorlet{tablebest}{tableaccent}
\colorlet{tablereference}{tableaccent}
\definecolor{tableyes}{HTML}{16845B}
\definecolor{tableno}{HTML}{C23B3B}
\definecolor{tablepartial}{HTML}{D17B0F}
\newcommand{\cmark}{\textcolor{tableyes}{\ding{51}}}
\newcommand{\xmark}{\textcolor{tableno}{\ding{55}}}
\newcommand{\pmark}{\textcolor{tablepartial}{\ensuremath{\boldsymbol{\sim}}}}
\colorlet{tableheader}{figurethreeblue!19}
\colorlet{tablegroupgray}{black!6}
\colorlet{tableour}{figurethreeblue!9}
\newcommand{\scalecell}[1]{\textbf{#1}}

\newcommand{\bestcell}[1]{\textbf{#1}}
\newcommand{\secondbest}[1]{\underline{#1}}
\newcommand{\tworowhead}[1]{\multirow[c]{2}{*}[-4pt]{\textbf{#1}}}

\AtBeginEnvironment{table}{\small}
\AtBeginEnvironment{longtable}{\small}

\newtcolorbox{promptbox}{
  enhanced,
  breakable,
  colback=promptbg!5,
  colframe=promptborder,
  boxrule=0.6pt,
  arc=2pt,
  left=8pt,
  right=8pt,
  top=8pt,
  bottom=8pt,
  before skip=8pt,
  after skip=10pt,
  fontupper=\small,
}

\title{\textbf{UA-Bench: Benchmarking Large Audio-Language Models on Underwater Acoustic Recognition}}

\author{%
\parbox{0.96\textwidth}{\centering
Zhe Sun,\,
Liang An,\,
Jiayu Yao,\,
Ronglai Qian,\,
Jianing Cai,\,
Ziwei Huang,\,
Qing Fan,\,
Zhiyuan Yang,\,
Yujun Cai,\,
Hongli Cao,\,
Yiwei Wang,\,
Xiaoyan Wang$^{\S}$\\
}}
\affiliation{%
\parbox{0.96\textwidth}{\centering
Southeast University\\
\email{zhe.sun@seu.edu.cn}\quad \email{xyan@seu.edu.cn}
}}

\abstract{
Large audio-language models (LALMs) are fast becoming a general interface to sound, yet how far their broad pretraining carries into specialized domains remains largely open. Underwater acoustics is a case in point. Vessel monitoring, maritime safety, and ocean sensing all depend on recognizing underwater signals, but to our knowledge LALMs have not been systematically evaluated on them. Building a benchmark for this setting is not straightforward. Annotated recordings are costly to collect, and a faithful synthetic alternative has to bring together many interacting factors: the structure of detection pulses, communication waveforms, and ship-radiated noise, and the sound-speed profile, geometry, and multipath propagation through which each reaches the receiver. We address this by proposing \textbf{UA-Bench}, to our knowledge the first large-scale benchmark for evaluating LALMs on underwater acoustic signal recognition. 
It comprises 375,051 audio records that couple parameterized sources with modeled propagation, so that every record can be traced to the conditions that produced it. The records span these three signal families, and each carries a three-level label hierarchy (13 leaf classes), a templated multi-turn dialogue, and source and channel metadata. We evaluate on a balanced 2,600-record subset with a gated coarse-to-fine protocol that locates each error at source routing, fine-grained classification, or output format, and add a paired silent-audio control to test whether the waveform is actually used. Nine open LALMs, evaluated without task-specific adaptation, reach only 2.0\%–13.1\% leaf accuracy, against a random reference of 7.7\%. The waveform helps source routing far more than leaf recognition (for one model, +29.5 versus +2.3 points over silence), and just four models gain reliably at the leaf level, by at most 4.7 points. UA-Bench thus gives a reference point for these models, and its released training partition supports follow-up adaptation work. 
}

\checkdata[Source Code and Data:]{\url{https://sunzhe12.github.io/ua-bench/}}

\begin{document}
\maketitle
\begingroup
\renewcommand{\thefootnote}{\fnsymbol{footnote}}
\footnotetext[4]{Corresponding author.}
\endgroup

\section{Introduction}
\label{sec:intro}

\begin{figure}[!t]
    \centering
    \includegraphics[width=\textwidth]{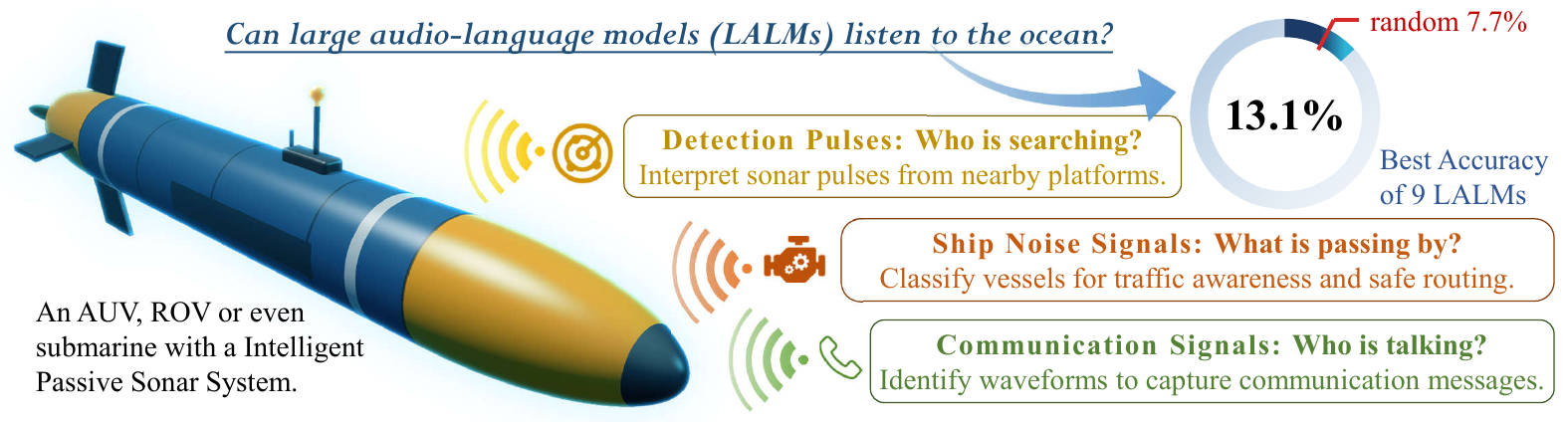}
    \caption{\textbf{UA-Bench at a glance.}
  An underwater platform (AUV, ROV, or submarine) equipped with an intelligent passive sonar system must interpret the sound field around it.
  We build UA-Bench, the first underwater acoustic understanding benchmark, to evaluate large audio-language models (LALMs) on this task.
  Despite their strong general-purpose audio understanding, the best of the nine evaluated LALMs reaches only 13.1\% accuracy at the finest level (L3), compared with 7.7\% for the random-guess reference, which shows that current LALMs are far from ready to listen to the ocean.}
    \label{fig:at_glance}
\end{figure}

Large audio-language models (LALMs) pair an audio encoder with a language model and are applied to a widening range of audio understanding tasks \cite{chu2023qwen,tang2024salmonn,gama}. Benchmarks for these models have centered on speech, music, and environmental sound \cite{yang2024air,sakshi2025mmau,huang2025dynamic}. Signals generated by physical or engineered processes have received less attention. Their categories are typically defined by temporal and spectral structure rather than by semantic content, so recognizing them may depend on cues that everyday sound categories seldom require. How LALMs perform on such signals remains largely unexamined.

Underwater acoustic recognition provides a concrete setting, as illustrated in Figure~\ref{fig:at_glance}. It supports vessel monitoring, maritime safety, and ocean observation \cite{shipsEar,deepShip,oceanship,doisy2008reverberation,stojanovic1996recent}, and its signals fall into three kinds, each with its own structure. Active detection pulses are distinguished by their frequency trajectories, communication waveforms by their modulation, and ship-radiated noise by its tonal and broadband composition, with considerable overlap among vessel categories. Multipath propagation, which depends on the sound-speed profile and on the source and receiver geometry, further reshapes each signal before it reaches the receiver \cite{stojanovic2009channels}. Labeled data are also scarce because annotated recordings are costly to acquire \cite{cui2024crossdomain}.

Existing resources address this setting only in part. General benchmarks such as AIR-Bench and MMAU cover a wide range of audio tasks, but they contain no engineered detection pulses or underwater communication classes and provide no controlled channel metadata~\cite{yang2024air,sakshi2025mmau}. Underwater datasets such as ShipsEar, DeepShip, QiandaoEar22, and Oceanship offer valuable field recordings, yet most support supervised, closed-set vessel recognition~\cite{shipsEar,deepShip,qiandaoear22,oceanship}. And the closest marine-domain example, WoW-Bench, relies on marine mammal vocalizations~\cite{kim2025wowbench}. What is needed is a resource that brings engineered underwater signals, known generation conditions, and a question-answering interface together in one evaluation.

Building such a resource and evaluating LALMs with it raise difficulties on two sides. The first lies in the data. A field recording reflects the source and the propagation path together, and the conditions behind it are seldom fully known, so an error is difficult to trace to either. Generating signals under known conditions can supply the missing control, but it involves many interacting factors. The three signal families follow different generative models, from fixed-carrier and swept pulses~\cite{doisy2008reverberation} to digitally modulated waveforms~\cite{stojanovic1996recent} and the tonal and broadband radiation of vessels~\cite{ross1976mechanics,wales2002ensemble}. The received waveform further depends on the sound-speed profile, the source and receiver geometry, and the multipath arrivals they produce~\cite{stojanovic2009channels}.

The second lies in the evaluation. Recognition proceeds from coarse choices, such as whether a signal is transmitted deliberately (active) or radiated by a platform (passive), to fine ones, such as the modulation scheme or the vessel type, and an overall accuracy does not show at which step a model breaks down. A model may also answer from the wording of a question and from class priors, so a correct answer does not by itself show that the recording was used~\cite{foo2025allthatglitters}.

\begin{figure}[!t]
    \centering
    \includegraphics[width=\textwidth]{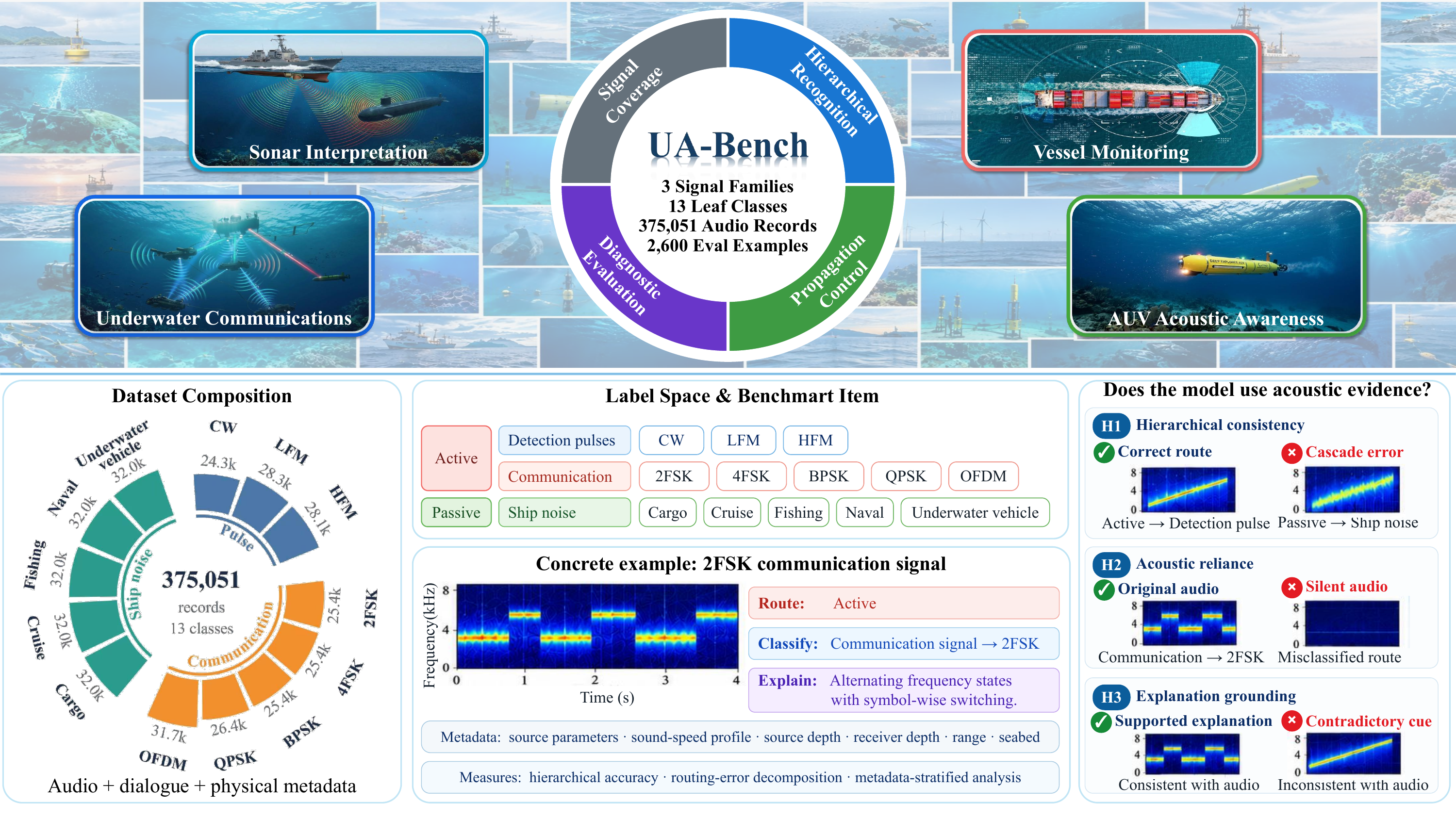}
    \caption{\textbf{Overview of UA-Bench.} Detection pulses, communication waveforms, and ship-radiated noise are synthesized under traceable source and propagation conditions. Each record includes a three-level label and a reference dialogue that proceeds from source category to leaf class and then requests an explanation. The benchmark measures recognition accuracy, separates routing errors from downstream errors, and compares performance across acoustic conditions. It also checks whether each explanation agrees with the predicted label.}
    \label{fig:overview}
\end{figure}

We introduce \textbf{UA-Bench}, a benchmark for evaluating LALMs on underwater acoustic signals, meant to complement field recordings (Figure~\ref{fig:overview}). It addresses the two sets of difficulties through its data and its evaluation protocol. To make errors attributable, each source is generated from a parameterized model of its class, covering three kinds of detection pulses, five communication modulations, and noise radiated by five vessel types, and is propagated with BELLHOP under sound-speed profiles drawn from ocean climatology and perturbed geometry~\cite{porter2019bellhop,woa2023}. The generating parameters are stored with every result, which yields 375,051 audio clips in 13 leaf classes, each traceable to the source and channel conditions that produced it and paired with a templated three-turn dialogue. Clips of the same source are kept in the same partition, so that propagation variants of one source cannot leak across partitions. A balanced subset of 2,600 clips, UA-Bench-Eval, is fixed for evaluation, and the training partition is released to support later adaptation studies.

To locate errors, evaluation follows a gated coarse-to-fine dialogue over a three-level label that runs from the source category through the signal family to the leaf class. A model names the source category, selects a complete class path within the chosen branch, and then explains its choice. A wrong answer at the first turn ends the dialogue, so every error is assigned to source routing (the active or passive decision), fine-grained classification, or output format. To test reliance on the waveform, each evaluation recording is paired with silence of equal duration under identical questions, and the two sets of answers are compared.

We evaluate nine open LALMs without task-specific adaptation. Leaf accuracy, the share of clips whose complete class path is correct, ranges from 2.0\% to 13.1\%, against a random reference of 7.7\%, and no model is best on all three signal families. The models fail in different ways: some at the answer format, some at source routing, and some by concentrating their answers on a few classes. In the silent control, the recording improves source routing for some models and changes it little for others, whereas the gain in leaf accuracy is at most about 4.7 points for any model. For the models that benefit, the contribution of the waveform appears mainly at the coarse level, a pattern that an overall accuracy would have concealed.

Our contributions are as follows.
\begin{itemize}
 \item \textbf{A controlled benchmark.} UA-Bench brings detection pulses, communication waveforms, and ship-radiated noise into one audio-language task, with hierarchical labels, templated dialogues, traceable source and channel metadata, and source-grouped partitions.
 \item \textbf{A diagnostic evaluation design.} A gated coarse-to-fine protocol, metadata-conditioned analysis, and a paired silent-audio control separate signal from channel, locate the failing decision, and test reliance on the waveform.
 \item \textbf{An empirical account of current LALMs.} Nine open models show low leaf accuracy, divergent failure modes, and limited use of the waveform at the leaf level.
 \end{itemize}

\FloatBarrier

\section{Related Work}
\label{sec:related}

Research related to UA-Bench spans general audio-language evaluation, the transfer of pretrained audio models to specialized acoustics, and underwater recognition with measured or simulated data. Table~\ref{tab:bench_compare} compares UA-Bench with representative work in these areas. Existing audio-language benchmarks rarely include underwater signals, while underwater resources usually evaluate task-specific models rather than LALMs. WoW-Bench is the closest marine-domain LALM benchmark and contains 1,777 validated questions about marine mammal vocalizations~\cite{kim2025wowbench}. UA-Bench instead covers three underwater signal families and 13 leaf classes in 375,051 records, together with a fixed evaluation suite of 2,600 records. It is the first large-scale, multi-category benchmark for evaluating LALMs on underwater acoustic signal recognition, with controlled source and propagation metadata for analysis under known conditions.

\begin{table}[!t]
\centering
\small
\setlength{\tabcolsep}{2.4pt}
\setlength{\aboverulesep}{0.15ex}
\setlength{\belowrulesep}{0.15ex}
\renewcommand{\arraystretch}{1.18}
\caption{\textbf{Comparison with representative audio-language benchmarks and underwater acoustic resources.}}
\label{tab:bench_compare}
\resizebox{\textwidth}{!}{%
\begin{tabular}{llcl!{\color{figurethreeblue!45}\vrule width 0.5pt}ccccc}
\toprule
\rowcolor{tableheader}
\textbf{Resource} & \textbf{Signal focus} & \textbf{Scale} & \textbf{Primary target} & \textbf{LALM} & \textbf{Hier.} & \textbf{Dialogue/QA} & \textbf{Source meta.} & \textbf{Prop. ctrl.} \\
\midrule
\rowcolor{tablegroupgray}
\multicolumn{9}{l}{\textbf{\textit{Audio-language benchmarks}}} \\
AIR-Bench~\cite{yang2024air} & Speech, sound, music & \scalecell{21k Q} & Found. + chat & \cmark & \xmark & \cmark & \xmark & \xmark \\
\rowcolor{figurethreeblue!3}
MMAU~\cite{sakshi2025mmau} & Speech, sound, music & \scalecell{10k QA} & Percept. + reason. & \cmark & \xmark & \cmark & \xmark & \xmark \\
Dynamic-SUPERB P2~\cite{huang2025dynamic} & Speech, music, sound & \scalecell{180 tasks} & Cls. / reg. / gen. & \pmark & \xmark & \pmark & \pmark & \xmark \\
\rowcolor{figurethreeblue!3}
AudioBench~\cite{wang2025audiobench} & Speech, voice, scenes & \scalecell{100k+ samp.} & 8 tasks / 26 sets & \cmark & \xmark & \pmark & \pmark & \xmark \\
WoW-Bench~\cite{kim2025wowbench} & Marine mammal calls & \scalecell{1,777 Q} & Percept. + cognit. & \cmark & \xmark & \cmark & \xmark & \xmark \\
\midrule
\rowcolor{tablegroupgray}
\multicolumn{9}{l}{\textbf{\textit{Underwater acoustic resources}}} \\
ShipsEar~\cite{shipsEar} & Recorded vessel noise & \scalecell{90 rec.} & 11 vessel types & \xmark & \xmark & \xmark & \cmark & \xmark \\
\rowcolor{figurethreeblue!3}
DeepShip~\cite{deepShip} & Recorded vessel noise & \scalecell{47 h} & 4 vessel classes & \xmark & \xmark & \xmark & \pmark & \xmark \\
QiandaoEar22~\cite{qiandaoear22} & Recorded vessel noise & \scalecell{10,611 rec.} & Detection + ID & \xmark & \xmark & \xmark & \cmark & \xmark \\
\rowcolor{figurethreeblue!3}
Oceanship~\cite{oceanship} & Recorded vessel noise & \scalecell{121 h} & 15 classes + retr. & \xmark & \xmark & \xmark & \cmark & \xmark \\
\midrule
\rowcolor{tableour}
\textbf{UA-Bench (ours)} & \textbf{Pulses, comms., vessel noise} & \scalecell{375,051 rec.} & \textbf{13-leaf hier. rec.} & \cmark & \cmark & \cmark & \cmark & \cmark \\
\bottomrule
\end{tabular}
}
\vspace{2pt}

\begin{minipage}{0.99\textwidth}
\scriptsize \cmark denotes supported, \pmark denotes partial or dataset-dependent support, and \xmark denotes no support. Q denotes questions, samp. denotes samples, rec. denotes recordings, and retr. denotes retrieval. ``Source meta.'' refers to source-condition metadata. ``Prop. ctrl.'' refers to independently specified propagation conditions.
\end{minipage}
\end{table}

\subsection{Audio Understanding Benchmarks}

General audio benchmarks measure capabilities across spoken content, instruments, sound events, perception, and reasoning. AIR-Bench~\cite{yang2024air} includes foundational tasks in speech, sound, and music, and it also supports open-ended interaction. MMAU~\cite{sakshi2025mmau} places greater emphasis on expert-level perception and reasoning across speech, environmental audio, and music. AudioBench~\cite{wang2025audiobench} evaluates speech, voice, and diverse acoustic scenes. Broader evaluation frameworks include AHELM~\cite{ahelm}, which combines several datasets and metrics, and VoiceBench~\cite{voicebench}, which focuses on voice assistant capabilities.

Recent work has moved beyond conventional sound-event recognition. Dynamic-SUPERB Phase-2~\cite{huang2025dynamic} extends instruction-based evaluation to classification, regression, and sequence generation, with some tasks examining acoustic properties such as pitch and rhythm. GAMA~\cite{gama} studies more complex audio reasoning through the synthetic CompA-R instruction-tuning dataset and the human-annotated CompA-R-test. WoW-Bench~\cite{kim2025wowbench} further extends fine-grained perception and cognition to marine mammal vocalizations. These studies broaden the scope of audio evaluation, but engineered underwater signals and variation caused by propagation remain largely outside it.

\subsection{General Audio Pretraining in Specialized Acoustics}

Pretrained audio representations are increasingly used in domains that require specialized acoustic evidence. BioLingual applies contrastive audio and language pretraining to more than one million captioned animal recordings, supporting zero-shot species recognition and natural-language retrieval across taxa~\cite{robinson2024biolingual}. General audio models have also been tested on respiratory and heart sounds, where frozen encoders provide features for downstream classifiers. Their performance can approach task-specific systems on clean recordings but declines on noisier tasks~\cite{niizumi2025audiofm}. These results demonstrate both the value and the limits of broad audio pretraining.

Most studies in specialized domains use a learned probe, fine-tuning, or further domain pretraining, so they do not show whether an instruction-following LALM can recognize the same signals without task-specific adaptation. Source structure and acquisition conditions are also rarely varied independently. UA-Bench focuses directly on this transfer setting by keeping the model fixed while varying the source family and propagation conditions in a controlled manner. The values of these variables are retained for diagnostic analysis.

\subsection{Underwater Acoustic Recognition, Data, and Simulation}

Underwater target recognition has traditionally relied on task-specific spectral and modulation features. Work on ShipsEar~\cite{shipsEar} combines cepstral representations with Gaussian mixture models for vessel classification. DeepShip~\cite{deepShip} compares time-domain and frequency-domain representations and evaluates several learning methods, including a separable convolutional autoencoder. QiandaoEar22~\cite{qiandaoear22} uses LOFAR and DEMON spectra to capture narrowband and modulation patterns for individual ship identification. Cross-domain contrastive learning has also been used to address limited annotations and shifts between recording environments~\cite{cui2024crossdomain}.

The datasets differ in annotation detail and task coverage. ShipsEar and DeepShip mainly provide vessel-category labels, while QiandaoEar22 supports the identification of individual targets. Oceanship~\cite{oceanship} adds vessel coordinates, velocity, and timestamps to its recordings and supports both classification and retrieval between audio and text. These resources preserve the complexity of real underwater environments, but the source and its propagation effects are observed together and cannot be controlled independently. It is therefore difficult to isolate the effect of a specific channel on a known source.

Physical simulation provides a complementary approach because it separates source generation from channel transformation. Classical ship-noise models describe the broadband, tonal, and modulation components of radiated waveforms~\cite{ross1976mechanics,arveson2000radiated,wales2002ensemble}. The Acoustics Toolbox provides several propagation solvers, including BELLHOP based on Gaussian beams and KRAKEN based on normal modes~\cite{porter1987gaussian,porter1984normal,porter2019bellhop}. Their inputs specify the sound-speed profile, boundaries, source and receiver depths, and range, which allows researchers to recompute arrivals or acoustic fields under different conditions.

These simulators are forward models rather than evaluation resources, and they do not define a source inventory, label space, leakage-safe partition, dialogue task, or diagnostic protocol. UA-Bench connects these elements in a single evaluation framework by combining parameterized source synthesis with BELLHOP propagation and validating each received waveform. It also retains the generating metadata and uses source-grouped data partitions, allowing model behavior to be related to known source and channel variables.

\FloatBarrier

\section{UA-Bench}
\label{sec:ua-bench}

UA-Bench is built to show where a LALM goes wrong when it listens to an underwater signal. It combines three components. A synthesis pipeline generates each source waveform from known parameters, propagates it through a simulated channel, and pairs the received audio, called a clip, with labels, a templated dialogue, and the source and channel conditions that produced it (Sections 3.2 and 3.3). A source-grouped partition, together with a balanced, fixed subset of the test partition, keeps training, validation, and evaluation data cleanly separated (Section 3.4). A gated coarse-to-fine dialogue assigns each error to a decision stage, and a paired silent-audio control tests whether the waveform contributes to a score (Sections 3.5 and 3.6).

\subsection{Task Scope and Label Hierarchy}
\label{sec:task}

\begin{table}[!htbp]
\centering
\small
\caption{\textbf{UA-Bench label hierarchy.} A source category, signal family, and specific class form each label path.}
\label{tab:taxonomy}
\begin{tabularx}{\textwidth}{@{}p{0.12\textwidth}p{0.22\textwidth}X@{}}
\toprule
\textbf{L1} & \textbf{L2} & \textbf{L3 classes} \\
\midrule
\rowcolor{tableaccent}[0pt][0pt]
\multicolumn{3}{@{}l@{}}{\textit{Active source category}} \\
Active & Detection pulse & CW, LFM, HFM \\
Active & Communication signal & 2FSK, 4FSK, BPSK, QPSK, OFDM \\
\rowcolor{tableaccent}[0pt][0pt]
\multicolumn{3}{@{}l@{}}{\textit{Passive source category}} \\
Passive & Ship-radiated noise & Cargo vessel, Cruise ship, Fishing vessel, Naval vessel, Underwater vehicle \\
\bottomrule
\end{tabularx}
\end{table}

UA-Bench poses recognition as a short dialogue about a single clip. The model
receives the clip's mono 16\,kHz waveform with the first question and keeps it
in the dialogue history, then answers three questions in turn. The first two
are multiple choice, answered with a single option letter: Turn~1 names the
source category, and Turn~2 names a complete class path. The third asks for a
brief explanation in free text. No task examples are given, and labels and
metadata are withheld from the model.

The answers follow a three-level hierarchy, which allows each failure to be
assigned to a coarse or a fine decision. Every example carries a label path
$y = (y^{(1)}, y^{(2)}, y^{(3)})$ for source category, signal family, and
specific class. At L1, \emph{Active} denotes signals transmitted for detection
or communication, and \emph{Passive} denotes noise radiated by an operating
vessel or underwater vehicle. At L2, the Active branch separates detection
pulses from communication waveforms, and the Passive branch contains
ship-radiated noise. These branches define the thirteen L3 classes in
Table~\ref{tab:taxonomy}. Turn~1 determines L1. Turn~2 lists the complete paths
of the branch reached at Turn~1, eight for Active and five for Passive, so a
single letter fixes L2 and L3 together. Classification and explanation are
scored separately (Section~\ref{sec:protocol}).

The classes are defined by the generator. Active classes follow waveform
structure. Vessel classes follow class-specific ranges of spectral level, shaft
frequency, blade count, tonal frequency, and modulation strength. The
underwater vehicle class, a generic propeller-driven platform, uses five or
seven blades, tonal frequencies of 10--200\,Hz, and principal modulation
coefficients of 0.06--0.08, while the surface-vessel classes use three to five
blades, tonal frequencies up to 1000\,Hz, and larger coefficients, with
overlapping ranges among them (Appendix~\ref{sec:appendix_synthesis}). UA-Bench therefore
measures recognition of generator-defined categories under known conditions,
not identification of real vessel types.

\subsection{Controlled Source and Channel Synthesis}
\label{sec:synthesis}

\begin{figure}[!t]
    \centering
    \includegraphics[width=\textwidth]{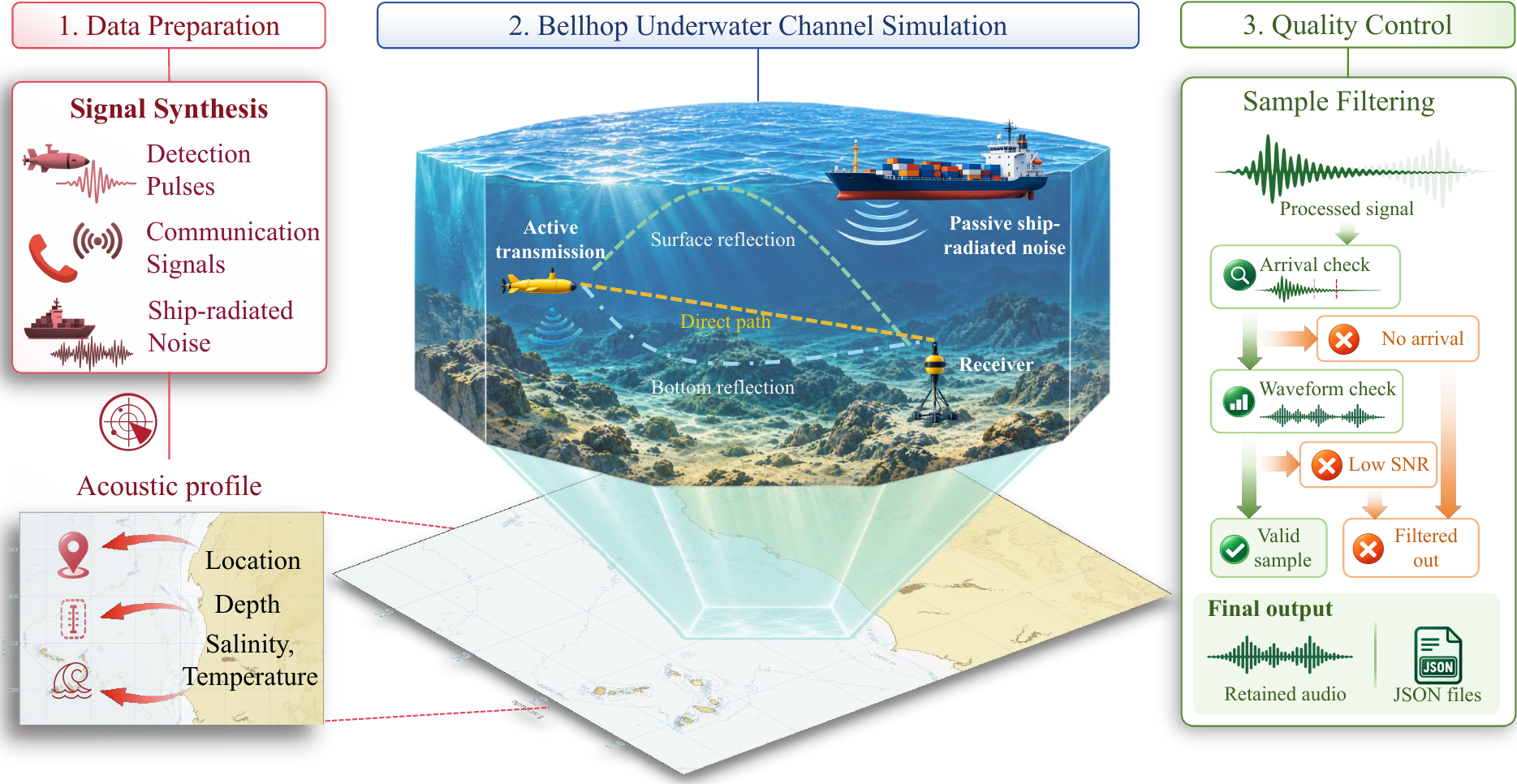}
    \caption{\textbf{Source generation and propagation in UA-Bench.} Parameterized underwater signals are propagated with BELLHOP under sampled environmental and geometric conditions. Arrival and waveform checks determine the retained audio and metadata.}
    \label{fig:channel_pipeline}
\end{figure}

To attribute a failure to signal structure or to propagation, the two have to
be generated separately. Each clip is therefore built in three steps: a
class-conditioned source, a sampled propagation channel, and output
normalization. For source~$i$ and propagation realization~$j$,
\begin{equation}
x_i=g_{y_i^{(3)}}(\theta_i),\qquad
u_{ij}=\mathcal H_{\phi_{ij}}[x_i],\qquad
a_{ij}=0.95\,\frac{u_{ij}}{\lVert u_{ij}\rVert_\infty},
\label{eq:generation}
\end{equation}
where $\theta_i$ holds the source parameters and $\phi_{ij}$ the channel parameters: the sound-speed profile, the source and receiver geometry, and the bottom properties (Figure~\ref{fig:channel_pipeline}).

\paragraph{Source generation.} All sources are synthesized at 16\,kHz. CW pulses use a fixed carrier, and LFM and HFM pulses follow linear and hyperbolic frequency trajectories. Communication waveforms use binary and quaternary frequency-shift keying, binary and quadrature phase-shift keying, and orthogonal frequency-division multiplexing~\cite{doisy2008reverberation,stojanovic1996recent}. Carrier and center frequencies are drawn from subbands between 300 and 7500\,Hz, while pulse width, sweep direction, symbol rate, symbol count, and modulation parameters vary across sources (Appendix~\ref{sec:appendix_synthesis}). A ship-radiated noise source combines a shaped continuous spectrum with discrete mechanical tones. The continuum is amplitude modulated at the shaft frequency, the blade-passing frequency, and one to four weaker modulation frequencies. Each source contains three to ten tones, of which one to three are tied to shaft or blade harmonics, and the ranges are class specific~\cite{ross1976mechanics,arveson2000radiated,wales2002ensemble}. Continuum and tones are mixed before propagation, so both undergo the same channel transformation.

\paragraph{Propagation.} Each realization is simulated with BELLHOP~\cite{porter2019bellhop}. The sound-speed profile is drawn from 100 annual-mean World Ocean Atlas profiles along a 15\,km transect in the northern South China Sea. Source depth, receiver depth, and range are perturbed independently by up to 10\%, and the seabed is silt in all simulations. Every source is propagated through up to two such realizations, each with its own deterministic random stream. BELLHOP returns the arrival amplitudes and delays, from which an impulse response is built (Appendix~\ref{sec:appendix_bellhop}). Pulse and communication signals are convolved with one impulse response, computed at the center or carrier frequency. Ship sources use responses at 50, 100, 200, 400, 800, and 1600\,Hz, interpolated across frequency. The convolution is cropped to the source duration.

\begin{figure}[!t]
\centering
\includegraphics[width=\textwidth]{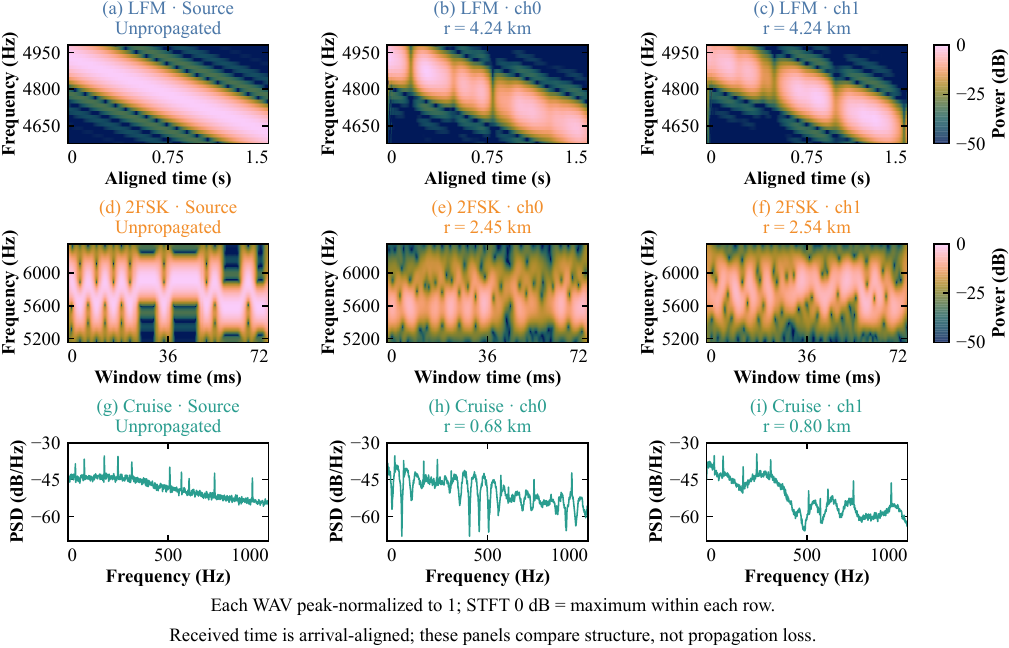}
\caption{\textbf{Representative source and channel realizations.} Each row follows one source waveform through two retained propagation realizations: (a--c) LFM, (d--f) 2FSK, and (g--i) cruise ship. The left column shows the unpropagated source; the middle and right columns show channel realizations \texttt{ch0} and \texttt{ch1}, with source-to-receiver range reported in each panel. Received time axes are aligned to the first arrival, and each waveform is peak-normalized independently; the comparison therefore concerns temporal and spectral structure rather than absolute propagation loss. Spectrogram power is referenced to the maximum within each row.}
\label{fig:channel_examples}
\end{figure}

\paragraph{Validation and normalization.} A realization is retained only if it yields at least one usable arrival and a finite, nonzero waveform. A simulation that returns no arrivals is repeated once with a denser beam grid. Each retained waveform is then scaled to a peak amplitude of 0.95. This removes absolute level and keeps relative temporal and spectral structure, so clips differ in structure rather than in propagation loss. The metadata record the source parameters, sampled environment, arrival information, delay reference, normalization gain, and duration.

Figure~\ref{fig:channel_examples} compares one source passed through two independent realizations. Multipath reshapes the local time--frequency pattern of the active signals and the spectral envelope of ship noise, while the class-defining structure stays recognizable.

\subsection{QA Annotation Construction}
\label{sec:qa_construction}

The synthesis stage produces clips with known label paths and source and channel conditions. We next pair each clip with questions and reference answers for the dialogue task defined in Section~\ref{sec:task}. The annotated dataset is
\begin{equation}
\begin{aligned}
\mathcal D &= \{(a_i,\mathbf y_i,\mathcal Q_i,\mathbf m_i,s_i)\}_{i=1}^{N},\\
\mathcal Q_i &= ((q_{i1},r_{i1}),(q_{i2},r_{i2}),(q_{i3},r_{i3})).
\end{aligned}
\label{eq:record}
\end{equation}
Here the $N$ retained clips are reindexed by $i$. The waveform is denoted by $a_i$, its three-level label path by $\mathbf y_i$, its source and channel metadata by $\mathbf m_i$, and its underlying source identifier by $s_i$. Each pair $(q_{it},r_{it})$ contains the question and reference answer for turn $t\in\{1,2,3\}$.

\paragraph{Classification questions.} The first two turns translate the label path into multiple-choice questions. Turn~1 asks for L1 using two source-category options. Turn~2 asks for L2 and L3 together by presenting eight complete paths on the Active branch or five on the Passive branch. Each question requires the uppercase letter of one option, and the reference letters are determined by $\mathbf y_i$. The annotation procedure samples from eight Turn~1 templates, eight Active Turn~2 templates, and six Passive Turn~2 templates. The option order is fixed across templates. 

\paragraph{Reference explanations.} The third turn links the class decision to acoustic structure by asking for a brief explanation of the preceding classification. Its reference answer is assembled from the source metadata and combines a source descriptor, a family descriptor, and three acoustic descriptors. Active-signal descriptors follow the generated waveform structure, with the LFM and HFM references also using the recorded sweep direction. Vessel descriptors use tonal extent, line count, harmonic tags, and modulation coefficients. Appendix~\ref{sec:appendix_qa} lists the templates.

\paragraph{Record assembly.} These annotations are packaged with each clip in a JSONL record containing its audio path, three-turn dialogue, canonical labels, source identity, and source and channel metadata. The annotation code maps every label to a valid three-level path; missing audio paths and unsupported label paths cause record construction to fail. At evaluation time, the model receives only the audio, stored questions, and earlier model answers. Ground-truth labels, reference answers, and metadata are withheld. The source identifier $s_i$ is retained so that propagation realizations of the same source remain together when the records are partitioned in Section~\ref{sec:selection}.

\subsection{Dataset Partitions and Evaluation Selection}
\label{sec:selection}

The annotation stage produces records that are ready for use in the dialogue task, but multiple records can share one source waveform. We therefore split records by source identity to keep propagation variants of the same source from crossing training, validation, and test partitions. We then select a fixed, class-balanced subset of the test partition, UA-Bench-Eval, so that all models are compared on the same inputs with equal support for each leaf class.

\paragraph{From sources to records.} A source is a waveform before propagation; a record contains one retained clip and its annotations. The source collection contains 240,000 waveforms: 80,000 detection pulses, 80,000 communication waveforms, and 80,000 ship-radiated noise waveforms. The pulse set contains 24,000 CW signals and 28,000 signals for each chirp class. Each communication and vessel class contains 16,000 sources. Two configured propagation attempts per source allow at most 480,000 source--channel realizations. After channel validation and annotation, UA-Bench contains 375,051 records. Each record represents one retained realization, so records that share a source must be grouped before splitting. Table~\ref{tab:data_flow} distinguishes the source waveforms, retained records, and selected evaluation examples.

\begin{table}[!htbp]
\centering
\small
\caption{\textbf{Dataset construction and partition sizes.} Counts distinguish configured source waveforms, propagation attempts, retained audio records, source-grouped partitions, and the fixed evaluation subset.}
\label{tab:data_flow}
\begin{tabularx}{\textwidth}{@{}p{0.22\textwidth}>{\raggedleft\arraybackslash}p{0.13\textwidth}X@{}}
\toprule
\textbf{Stage} & \textbf{Count} & \textbf{Unit} \\
\midrule
\rowcolor{tableaccent}[0pt][0pt]
\multicolumn{3}{@{}l@{}}{\textit{Source generation and propagation}} \\
Configured sources & 240,000 & Waveforms before propagation \\
Propagation attempts & 480,000 & Maximum at two realizations per source \\
\rowcolor{tableaccent}[0pt][0pt]
\multicolumn{3}{@{}l@{}}{\textit{Released source-grouped records}} \\
Retained records & 375,051 & Audio, labels, dialogue, and metadata \\
Training & 262,580 & Records in training source groups \\
Validation & 56,250 & Records in validation source groups \\
Test & 56,221 & Records in test source groups \\
\textbf{UA-Bench-Eval} & \textbf{2,600} & Selected test records, 200 per class \\
\bottomrule
\end{tabularx}
\end{table}

\paragraph{Source-grouped splitting.} All records with the same source identifier $s_i$ form one group and remain in the same partition. Within each L3 class, these groups are shuffled with seed 42 and allocated to training, validation, and test partitions in a nominal ratio of 70, 15, and 15 percent. The largest-remainder rule converts the ratios to integer group counts. Every record inherits the partition of its source. The resulting record counts are shown in Table~\ref{tab:data_flow}, and UA-Bench-Eval is drawn only from the test partition.

\paragraph{Class-specific ranking.} To construct the evaluation subset, we rank test records within each leaf class using a statistic from the generation metadata. The ranking variable depends on the source branch. Active signals use the impulse-response energy gain $G_h$, computed before output normalization. Ship-radiated noise uses the source tonal-to-continuum power ratio $S_{\mathrm{src}}$, computed before propagation:
\begin{equation}
G_h=10\log_{10}\sum_n |h[n]|^2,\qquad
S_{\mathrm{src}}=10\log_{10}
\frac{\operatorname{mean}_n x_{\mathrm{tonal}}[n]^2}
{\operatorname{mean}_n x_{\mathrm{continuum}}[n]^2}.
\label{eq:quality}
\end{equation}
Here $h[n]$ is the active-signal channel impulse response, $x_{\mathrm{tonal}}[n]$ is the complete tonal component of a ship source, and $x_{\mathrm{continuum}}[n]$ is its modulated continuous component; $n$ indexes discrete-time samples. The ship statistic compares the mean-square powers of these two complete source components. It differs from the local line-to-continuum ratio used to set each tone during synthesis. Appendix~\ref{sec:metadata_provenance} defines both quantities.

\paragraph{Fixed evaluation subset.} Within each leaf class, test records are ordered by decreasing $G_h$ or $S_{\mathrm{src}}$, with sample identifiers resolving ties. The first 400 records form a ranked candidate pool. We sample 200 records without replacement from each pool with seed 42. This gives 2,600 evaluation records with equal support across the thirteen leaf classes: 600 detection-pulse records, 1,000 communication-signal records, and 1,000 ship-radiated noise records. The selection code verifies record identifiers, label paths, metadata, audio paths, and per-class counts.

\begin{figure}[!t]
\centering
\includegraphics[width=\textwidth]{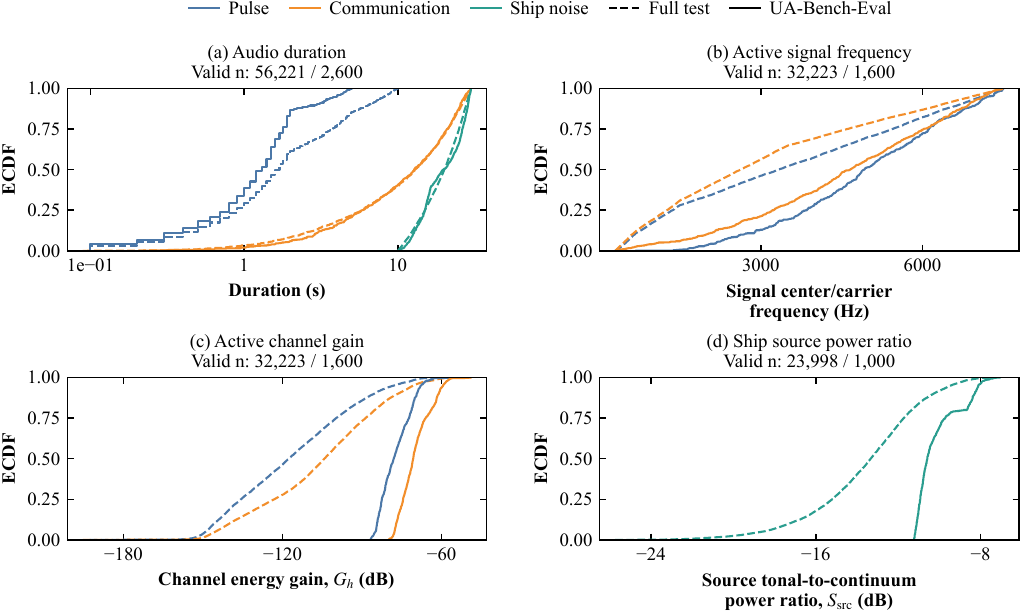}
\caption{\textbf{Distributions in the full test partition and UA-Bench-Eval.} Empirical cumulative distribution functions compare (a) audio duration, (b) active-signal center or carrier frequency, (c) active channel energy gain $G_h$, and (d) ship source tonal-to-continuum power ratio $S_{\mathrm{src}}$. Colors identify signal families; dashed and solid curves denote the full test partition and UA-Bench-Eval, respectively. Valid record counts are reported as full test / UA-Bench-Eval.}
\label{fig:selection_distributions}
\end{figure}

\paragraph{Properties of the selected inputs.} Ranking and sampling determine both the class counts and the acoustic conditions represented in UA-Bench-Eval. Figure~\ref{fig:selection_distributions} compares the selected records with the full test partition. The selected $G_h$ values for active signals and $S_{\mathrm{src}}$ values for ship noise are concentrated at larger values, and active center or carrier frequencies shift upward. Duration changes depend on the signal family: selected pulses are more concentrated at shorter durations, while the communication and ship-noise curves remain closer to their full-test counterparts.

The appendix galleries show the acoustic structures present in these evaluation inputs. Appendix Figure~\ref{fig:acoustic_active} shows stationary, linear, and curved pulse trajectories, together with discrete frequency switching and multicarrier occupancy. The power-only views of BPSK and QPSK are not interpreted as direct measurements of their phase states. Appendix Figure~\ref{fig:acoustic_ships} shows low-frequency tonal and continuous ship-noise structure. The displayed spectral-centroid intervals overlap strongly across the five ship-noise leaf classes.

With the evaluation inputs fixed, Section~\ref{sec:protocol} defines how their stored dialogues are executed and how the classification decisions are scored.

\subsection{Gated Evaluation Protocol and Classification Metrics}
\label{sec:protocol}

Each model receives the audio and stored questions from UA-Bench-Eval without task examples in the prompt. The first two turns yield predictions at the three label levels; the third supplies an explanation for separate analysis.

\paragraph{Gated execution.} At Turn~1, the evaluator compares the predicted source category with L1. An invalid-format or incorrect answer ends the dialogue and assigns \texttt{cascade\_error} to L2 and L3. A correct answer opens the corresponding branch at Turn~2, where one complete-path option determines both L2 and L3. Turn~3 is executed for every example that passes Turn~1, including examples with an invalid-format or incorrect Turn~2 answer.

\paragraph{Answer decoding.} For the two classification turns, the decoder removes surrounding whitespace and accepts only a single ASCII letter that matches a displayed option, treating lowercase and uppercase forms equivalently. Free-form text and expanded class names are not recovered. The raw answer and decoding status are preserved for error analysis.

Appendix Figure~\ref{fig:ua_bench_three_turn_dialogue} illustrates a 2FSK dialogue. Correct source routing opens the Active branch, where option D identifies the Communication family and the 2FSK leaf; Turn~3 then requests an acoustic explanation.

\paragraph{Accuracy and conditional accuracy.} For this evaluation, $N=2{,}600$ and $i$ indexes the records in UA-Bench-Eval. Let $c_i^{(\ell)}$ be one when the prediction at label level $\ell\in\{1,2,3\}$ matches its reference label and zero otherwise. Unresolved and cascade outcomes count as incorrect. Define $z_i^{(\ell)}=\prod_{k=1}^{\ell}c_i^{(k)}$ to indicate that the full path through level $\ell$ is correct. The metrics are
\begin{align}
\mathrm{Acc}_{L\ell}&=\frac{1}{N}\sum_{i=1}^{N}c_i^{(\ell)},\label{eq:accuracy}\\
\mathrm{Acc}(L\ell\mid\mathrm{parent\ path})&=
\frac{\sum_i z_i^{(\ell)}}{\sum_i z_i^{(\ell-1)}}\qquad (\ell=2,3).
\label{eq:conditional}
\end{align}
We abbreviate the conditional quantities as L2$|$L1 and L3$|$L2. The first conditions on correct L1; the second conditions on correct L1 and L2. An empty conditional denominator yields a score of zero, following the convention in Appendix~\ref{sec:appendix_metrics}.

\paragraph{Class balance and failure outcomes.} L3 Macro F1 averages the F1 scores of the thirteen leaf classes. Per-class precision, recall, and confusion matrices summarize class-specific errors. Invalid-format and cascade outcomes contribute false negatives to their true classes and appear as separate prediction columns; they are not additional classes in the Macro F1 average. For stage-wise analysis, we partition records into five mutually exclusive outcomes: invalid-format or valid-but-wrong answers at Turn~1, the same two outcomes at Turn~2 after a correct route, and correct L3 decisions. Under the execution gate and strict decoder, correct L3 implies correct L1 and L2, so a separate joint score would duplicate L3 accuracy. Appendix~\ref{sec:appendix_metrics} gives the numerical conventions.

\paragraph{Random reference.} The random reference chooses L1 uniformly and then chooses a leaf uniformly from the selected branch. The balanced evaluation set has eight Active and five Passive leaf classes, so its expected leaf accuracy is
\begin{equation}
\mathbb E[\mathrm{Acc}_{L3}]
=\frac{8}{13}\frac{1}{2}\frac{1}{8}
+\frac{5}{13}\frac{1}{2}\frac{1}{5}
=\frac{1}{13}\simeq7.69\%.
\label{eq:random}
\end{equation}
Including the probability of choosing the correct source category, the expected leaf accuracies are 6.25\% on Active examples and 10\% on Passive examples. A predictor that always selects Active has an L1 accuracy of $8/13$, or 61.54\%. Appendix~\ref{sec:appendix_metrics} derives the family-level and conditional references.

\subsection{Diagnostic Analyses}
\label{sec:diagnostic_analyses}

We supplement classification scores with a paired audio control, analysis across generation conditions, and a check of explanation terminology.

\paragraph{Silent-audio control.} Each evaluation record is paired with an equal-duration silent waveform while its questions and labels remain fixed. We compare L3 accuracy and Macro F1 across the two conditions and report original-minus-silent differences. We obtain 95\% percentile intervals for the scores and differences by bootstrapping source clusters, keeping channel variants of each source together. Appendix~\ref{sec:appendix_metrics} gives the resampling settings and additional interval checks.

\paragraph{Metadata-conditioned performance.} Within each leaf class, we divide examples into low, middle, and high strata using empirical tertiles of $G_h$ for active signals or $S_{\mathrm{src}}$ for ship-radiated noise. The thresholds are fixed across models. For each stratum, we report L3 recall averaged with equal leaf weight within each branch. Appendix~\ref{sec:appendix_repro} gives the quantile conventions.

\paragraph{Lexical explanation consistency.} Each executed Turn~3 response is matched against a supporting set of terms compatible with the predicted leaf and a comparison set containing the remaining inventory terms. Shared terms support every compatible leaf. The retained matches define four categories: \emph{Aligned} for supporting terms only, \emph{Concept Confusion} for comparison terms only, \emph{Vague} for neither, and \emph{Term Stacking} for both. The contradiction rate is the sum of the confusion and stacking rates. All rates use the number of executed Turn~3 responses for that model as the denominator. Appendix~\ref{sec:appendix_terms} specifies the inventory, matching rules, and treatment of invalid outputs and empty responses. The rates describe consistency with the predicted class at the level of terminology.
\section{Experiments}
\label{sec:experiments}

We evaluate recognition without task examples across detection pulses, communication waveforms, and ship-radiated noise. The analysis begins with performance over the complete label hierarchy. It then examines source families, decision-stage errors, metadata strata, and explanation vocabulary. This order connects the aggregate scores with the predictions from which they are derived.

\subsection{Experimental Setup}

\paragraph{Models and inference.} We evaluate nine released instruction-following checkpoints. They are Gemma-4-12B-IT and Gemma-4-E2B-IT~\cite{gemma4}, Audio-Flamingo-Next~\cite{audioflamingonext}, MiDashengLM-7B-1021~\cite{midashenglm}, Voxtral-Small-24B-2507~\cite{voxtral_small}, Voxtral-Mini-3B-2507~\cite{voxtral_mini}, Aero-1-Audio~\cite{aero1audio}, Qwen2.5-Omni-7B~\cite{qwen25omni}, and Qwen2-Audio-7B-Instruct~\cite{chu2024qwen2}. Each model uses its native audio processor and chat template. The audio is attached to the first user turn and retained in the dialogue history. Text generation uses greedy decoding when the backend supports it. Each turn permits at most 256 new tokens, and audio generation is disabled. Appendix~\ref{sec:appendix_repro} lists the checkpoints and settings.

\paragraph{Evaluation records.} All results use the fixed UA-Bench-Eval suite defined in Section~\ref{sec:selection}. The training and validation partitions are not used for model evaluation. All nine checkpoints have complete, integrity-verified strict-decoder runs and paired silent-audio controls. Run-specific batch sizes and software environments are preserved in the protocol records, so the reported comparisons should be interpreted as checkpoint-configuration results rather than architecture-only effects.

\subsection{Overall and Family Performance}
\label{sec:overall}

Table~\ref{tab:hierarchical_results} reports accuracy at all three levels, L3 Macro F1, and two conditional accuracies. The level accuracies measure predictions at individual levels. The conditional scores separate source routing from classification within a correctly reached path. Macro F1 measures the balance of terminal predictions across the thirteen leaves. A separate joint column is omitted because, under the strict gated decoder, a correct L3 decision necessarily identifies the complete path.

\begin{table}[!htbp]
\centering
\small
\setlength{\tabcolsep}{3.5pt}
\caption{\textbf{Hierarchical classification on 2,600 examples without task demonstrations.} All rows use the strict single-letter decoder. Accuracy values are percentages, and Macro F1 is reported on a scale from 0 to 1. Bold values mark the best model result, and underlining marks the second-best result. The shaded columns are the primary terminal-level metrics.}
\label{tab:hierarchical_results}
\begin{tabularx}{\textwidth}{@{}>{\raggedright\arraybackslash}X*{2}{>{\centering\arraybackslash}p{0.09\textwidth}}>{\columncolor{metricaccent}[\tabcolsep][\tabcolsep]\centering\arraybackslash}p{0.09\textwidth}>{\columncolor{metricaccent}[\tabcolsep][0pt]\centering\arraybackslash}p{0.11\textwidth}*{2}{>{\centering\arraybackslash}p{0.10\textwidth}}@{}}
\toprule
\tworowhead{Model} & \multicolumn{4}{c}{\textbf{Classification quality}} & \multicolumn{2}{c}{\textbf{Conditional accuracy}} \\
\cmidrule(lr){2-5}\cmidrule(lr){6-7}
& \textbf{L1 Acc.$\uparrow$} & \textbf{L2 Acc.$\uparrow$} & \textbf{L3 Acc.$\uparrow$} & \textbf{Macro F1$\uparrow$} & \textbf{L2$|$L1$\uparrow$} & \textbf{L3$|$L2$\uparrow$} \\
\midrule
Gemma-4-12B-IT~\cite{gemma4} & 27.6 & 9.3 & 2.0 & 0.023 & 33.9 & 21.0 \\
Gemma-4-E2B-IT~\cite{gemma4} & 51.6 & 30.7 & 6.5 & 0.049 & 59.5 & 21.2 \\
Audio-Flamingo-Next~\cite{audioflamingonext} & \bestcell{87.6} & 40.3 & 8.1 & 0.061 & 46.0 & 20.0 \\
MiDashengLM-7B-1021~\cite{midashenglm} & 57.5 & \secondbest{47.0} & \secondbest{10.3} & \bestcell{0.083} & \secondbest{81.7} & 21.9 \\
Voxtral-Small-24B-2507~\cite{voxtral_small} & 56.8 & 26.9 & 7.7 & 0.056 & 47.4 & \bestcell{28.7} \\
Voxtral-Mini-3B-2507~\cite{voxtral_mini} & 59.1 & 25.1 & 6.1 & 0.046 & 42.5 & 24.2 \\
Aero-1-Audio~\cite{aero1audio} & 61.3 & 24.3 & 6.9 & 0.033 & 39.6 & \secondbest{28.3} \\
Qwen2.5-Omni-7B~\cite{qwen25omni} & \secondbest{72.5} & \bestcell{54.3} & \bestcell{13.1} & \secondbest{0.082} & 74.8 & 24.2 \\
Qwen2-Audio-7B-Instruct~\cite{chu2024qwen2} & 38.5 & 38.5 & 7.3 & 0.063 & \bestcell{100.0} & 19.0 \\
\midrule
\textit{Gated random reference} & 50.0 & 35.6 & 7.7 & N/A & 71.2 & 21.6 \\
\bottomrule
\end{tabularx}
\end{table}
\FloatBarrier

Across the nine strict runs, L3 accuracy ranges from 2.0\% to 13.1\%. Qwen2.5-Omni obtains the highest accuracy, 5.4 percentage points above the gated random reference after rounding. MiDashengLM has the highest Macro F1 at 0.083, narrowly ahead of Qwen2.5-Omni at 0.082. Gemma-4-12B has the lowest L3 accuracy; its result is partly shaped by 1,156 Turn~1 refusals stating that no audio was available, even though the corresponding inputs passed the dataset and pipeline checks.

Earlier levels produce different rankings. Audio-Flamingo-Next has the highest L1 accuracy at 87.6\%, whereas Qwen2.5-Omni leads at L2 with 54.3\%. Qwen2-Audio attains 100.0\% L2$|$L1 because every correctly routed example also receives the correct family label, but its L1 accuracy is only 38.5\%. Voxtral-Small instead has the highest L3$|$L2 value at 28.7\%, narrowly followed by Aero-1-Audio at 28.3\%. These differences show why reporting the intermediate routing metrics remains useful even though L3 accuracy already represents complete-path correctness under the strict protocol.

Table~\ref{tab:family_results} separates terminal accuracy across the three source families. Detection pulses, communication waveforms, and ship-radiated noise contain 600, 1,000, and 1,000 examples. Their different best-performing models reveal that aggregate scores combine distinct acoustic strengths.

\begin{figure}[!t]
\centering
\includegraphics[width=\textwidth]{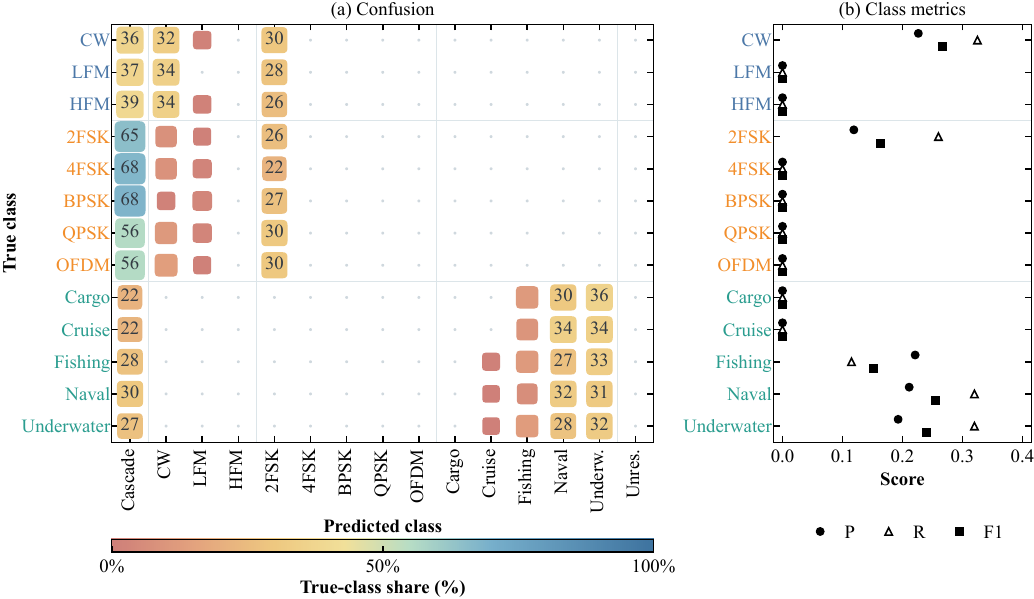}
\caption{\textbf{Class-level diagnostics for the strict MiDashengLM-7B-1021 run on 2,600 examples.} (a) Row-normalized confusion matrix over the thirteen leaf classes, with separate columns for cascaded (\emph{Cascade}) and unresolved (\emph{Unres.}) predictions. Colors encode true-class shares; values of at least 20\% are annotated as percentages rounded to integers. (b) Per-class precision (filled circles), recall (open triangles), and F1 (filled squares). The shortened labels \emph{Cargo}, \emph{Cruise}, \emph{Fishing}, and \emph{Naval} denote the corresponding taxonomy classes; both \emph{Underwater} and \emph{Underw.} denote \emph{Underwater vehicle}.}
\label{fig:class_diagnostics}
\end{figure}

\begin{table}[!t]
\centering
\small
\caption{\textbf{L3 accuracy by source family under the strict single-letter decoder.} Bold values mark the best model result, and underlining marks the second-best result.}
\label{tab:family_results}
\begin{tabular*}{\textwidth}{@{\extracolsep{\fill}}lccc@{}}
\toprule
\tworowhead{Model} & \multicolumn{3}{c}{\textbf{L3 accuracy by source family}} \\
\cmidrule(lr){2-4}
& \textbf{\shortstack{Detection pulse\\3 classes, $n=600$}} & \textbf{\shortstack{Communication waveform\\5 classes, $n=1{,}000$}} & \textbf{\shortstack{Ship-radiated noise\\5 classes, $n=1{,}000$}} \\
\midrule
Gemma-4-12B-IT~\cite{gemma4} & 4.5 & 1.0 & 1.4 \\
Gemma-4-E2B-IT~\cite{gemma4} & 4.3 & \bestcell{12.0} & 2.3 \\
Audio-Flamingo-Next~\cite{audioflamingonext} & 0.0 & \secondbest{6.2} & 14.8 \\
MiDashengLM-7B-1021~\cite{midashenglm} & 10.8 & 5.2 & 15.1 \\
Voxtral-Small-24B-2507~\cite{voxtral_small} & \bestcell{23.2} & 1.4 & 4.8 \\
Voxtral-Mini-3B-2507~\cite{voxtral_mini} & 16.8 & 2.2 & 3.5 \\
Aero-1-Audio~\cite{aero1audio} & \secondbest{22.5} & 4.4 & 0.0 \\
Qwen2.5-Omni-7B~\cite{qwen25omni} & 15.0 & 5.3 & \bestcell{19.8} \\
Qwen2-Audio-7B-Instruct~\cite{chu2024qwen2} & 0.0 & 0.0 & \secondbest{19.0} \\
\midrule
\textit{Gated random reference} & 6.25 & 6.25 & 10.0 \\
\bottomrule
\end{tabular*}
\end{table}
\FloatBarrier

Voxtral-Small has the highest result on detection pulses, Gemma-4-E2B-IT on communication waveforms, and Qwen2.5-Omni on ship-radiated noise. None of these models maintains its lead across the other two families. The strict runs are also highly concentrated: Aero-1-Audio reaches 61.5\% recall on CW but zero recall on every ship-noise leaf, Qwen2.5-Omni reaches 91.5\% on Naval vessel, and Qwen2-Audio obtains all of its correct terminal predictions on ship-noise examples.

The family averages also conceal substantial variation among leaf classes. Figure~\ref{fig:class_diagnostics} illustrates this variation for the strict MiDashengLM-7B-1021 run. Correct L3 predictions occur only for CW, 2FSK, Fishing vessel, Naval vessel, and Underwater vehicle; the remaining eight classes have zero recall. Cascade outcomes account for 53.0\% of active-signal examples and 25.7\% of ship-noise examples. Among correctly routed examples, active-signal predictions concentrate on CW and 2FSK, while ship-noise predictions concentrate on Naval vessel and Underwater vehicle. These results reveal both source-routing failures and limited discrimination among leaf classes within the correctly reached branches.

\begin{figure}[!t]
\centering
\includegraphics[width=\textwidth]{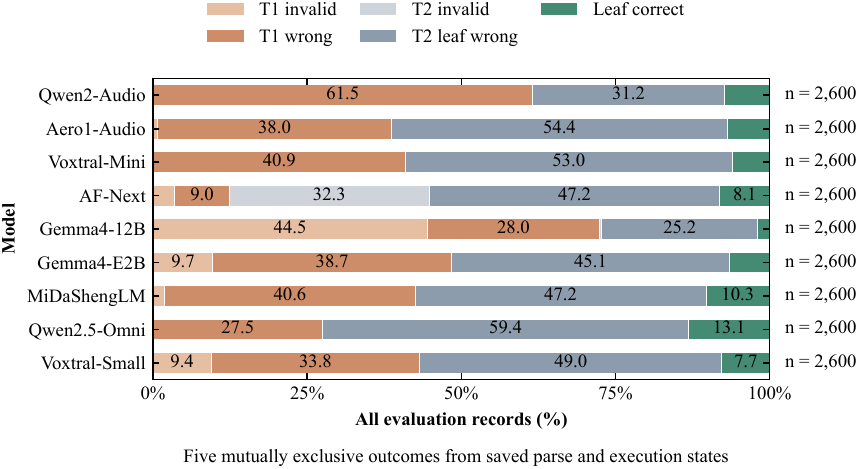}
\caption{\textbf{Decision-stage error decomposition under the strict single-letter decoder.} Each bar partitions 2,600 examples into invalid-format and valid-but-incorrect outcomes at Turn~1, the corresponding outcomes at Turn~2 after a correct route, and correct L3 decisions. The five segments reproduce the counts in Table~\ref{tab:failure_counts} as percentages.}
\label{fig:error_decomposition}
\end{figure}

\begin{table}[!t]
\centering
\small
\setlength{\tabcolsep}{4pt}
\caption{\textbf{Decision-stage outcomes under the strict single-letter protocol.} Counts sum to 2,600 in every row. The shaded column is the terminal success outcome.}
\label{tab:failure_counts}
\begin{tabularx}{\textwidth}{@{}>{\raggedright\arraybackslash}X*{4}{>{\centering\arraybackslash}p{0.13\textwidth}}>{\columncolor{metricaccent}[0pt][0pt]\centering\arraybackslash}p{0.11\textwidth}@{}}
\toprule
\tworowhead{Model} & \multicolumn{2}{c}{\textbf{Turn 1 failure}} & \multicolumn{2}{c}{\textbf{Turn 2 failure}} & \multicolumn{1}{c}{\textbf{Success}} \\
\cmidrule(lr){2-3}\cmidrule(lr){4-5}\cmidrule(lr){6-6}
& \textbf{Invalid format} & \textbf{Valid but wrong} & \textbf{Invalid format} & \textbf{Valid but wrong} & \textbf{L3 correct} \\
\midrule
Gemma-4-12B-IT~\cite{gemma4} & 1156 & 727 & 10 & 656 & 51 \\
Gemma-4-E2B-IT~\cite{gemma4} & 252 & 1006 & 0 & 1173 & 169 \\
Audio-Flamingo-Next~\cite{audioflamingonext} & 90 & 233 & 841 & 1226 & 210 \\
MiDashengLM-7B-1021~\cite{midashenglm} & 49 & 1056 & 0 & 1227 & 268 \\
Voxtral-Small-24B-2507~\cite{voxtral_small} & 245 & 879 & 0 & 1275 & 201 \\
Voxtral-Mini-3B-2507~\cite{voxtral_mini} & 0 & 1064 & 0 & 1378 & 158 \\
Aero-1-Audio~\cite{aero1audio} & 18 & 988 & 0 & 1415 & 179 \\
Qwen2.5-Omni-7B~\cite{qwen25omni} & 0 & 714 & 0 & 1545 & 341 \\
Qwen2-Audio-7B-Instruct~\cite{chu2024qwen2} & 0 & 1600 & 0 & 810 & 190 \\
\bottomrule
\end{tabularx}
\end{table}

The strict decoder distinguishes invalid-format answers from valid but incorrect decisions. Table~\ref{tab:failure_counts} partitions each run into five outcomes: two at source routing, two at terminal classification after a correct route, and a correct terminal decision. Because every valid Turn~2 option identifies a complete path, no separate leaf-only category is needed.

The strict-run error composition varies substantially. Gemma-4-12B produces 1,156 invalid-format Turn~1 answers, all of which are no-audio refusals, and only 51 correct terminal predictions. Audio-Flamingo-Next instead reaches Turn~2 more often but produces 841 invalid-format answers at that stage. Qwen2.5-Omni reaches Turn~2 on 1,886 examples without an invalid-format decision and produces 341 correct terminal predictions. Voxtral-Mini also produces no invalid-format answer, yet 1,378 examples are assigned a valid but incorrect terminal option. These differences illustrate why formatting failures and valid but incorrect decisions must be reported separately.

Figure~\ref{fig:error_decomposition} visualizes the five strict-decoder outcomes in Table~\ref{tab:failure_counts}. It shows that the dominant failure mode differs across checkpoints: Gemma-4-12B is dominated by Turn~1 invalid-format responses, Audio-Flamingo-Next has a substantial Turn~2 invalid-format segment, and several other models primarily produce valid but incorrect decisions. The terminal-success segment remains small for every model.

\subsection{Associations with Generation Metadata}
\label{sec:physical_results}

Table~\ref{tab:metadata_results} reports the metadata-stratified analysis defined in Section~\ref{sec:diagnostic_analyses}. Active records are grouped by $G_h$, while ship-radiated noise records are grouped by $S_{\mathrm{src}}$. Thresholds are computed within each leaf class and shared across models. Recall is first computed for each leaf and stratum. It is then averaged with equal leaf weight within each branch. The final column in each group reports the high-minus-low difference.

\begin{table}[!htbp]
\centering
\small
\setlength{\tabcolsep}{3.5pt}
\caption{\textbf{L3 recall across within-class metadata strata under the strict single-letter decoder.} Recall values are percentages. $\Delta$ is the high-minus-low difference in percentage points. The shaded columns show the high-minus-low effects.}
\label{tab:metadata_results}
\begin{tabularx}{\textwidth}{@{}>{\raggedright\arraybackslash}X*{3}{>{\centering\arraybackslash}p{0.067\textwidth}}>{\columncolor{metricaccent}[\tabcolsep][0pt]\centering\arraybackslash}p{0.067\textwidth}*{3}{>{\centering\arraybackslash}p{0.067\textwidth}}>{\columncolor{metricaccent}[\tabcolsep][0pt]\centering\arraybackslash}p{0.067\textwidth}@{}}
\toprule
\tworowhead{Model} & \multicolumn{4}{c}{\textbf{Active signals using $G_h$}} & \multicolumn{4}{c}{\textbf{Ship-radiated noise using $S_{\mathrm{src}}$}} \\
\cmidrule(lr){2-5}\cmidrule(lr){6-9}
& \textbf{Low} & \textbf{Middle} & \textbf{High} & \textbf{$\Delta$} & \textbf{Low} & \textbf{Middle} & \textbf{High} & \textbf{$\Delta$} \\
\midrule
Gemma-4-12B-IT~\cite{gemma4} & 1.5 & 1.7 & 3.7 & $+2.2$ & 0.9 & 2.1 & 1.2 & $+0.3$ \\
Gemma-4-E2B-IT~\cite{gemma4} & 6.9 & 10.4 & 10.1 & $+3.2$ & 1.5 & 3.0 & 2.4 & $+0.9$ \\
Audio-Flamingo-Next~\cite{audioflamingonext} & 4.3 & 3.6 & 3.7 & $-0.6$ & 14.3 & 14.5 & 15.5 & $+1.2$ \\
MiDashengLM-7B-1021~\cite{midashenglm} & 6.7 & 6.4 & 8.8 & $+2.1$ & 15.5 & 14.5 & 15.2 & $-0.3$ \\
Voxtral-Small-24B-2507~\cite{voxtral_small} & 9.5 & 9.7 & 9.5 & $0.0$ & 5.1 & 3.6 & 5.7 & $+0.6$ \\
Voxtral-Mini-3B-2507~\cite{voxtral_mini} & 8.4 & 9.3 & 5.4 & $-3.0$ & 3.6 & 4.2 & 2.7 & $-0.9$ \\
Aero-1-Audio~\cite{aero1audio} & 10.6 & 11.0 & 11.9 & $+1.3$ & 0.0 & 0.0 & 0.0 & $0.0$ \\
Qwen2.5-Omni-7B~\cite{qwen25omni} & 9.3 & 8.7 & 8.8 & $-0.6$ & 19.7 & 19.4 & 20.3 & $+0.6$ \\
Qwen2-Audio-7B-Instruct~\cite{chu2024qwen2} & 0.0 & 0.0 & 0.0 & $0.0$ & 16.4 & 19.4 & 21.2 & $+4.8$ \\
\bottomrule
\end{tabularx}
\end{table}
\FloatBarrier

\begin{figure}[!t]
\centering
\includegraphics[width=\textwidth]{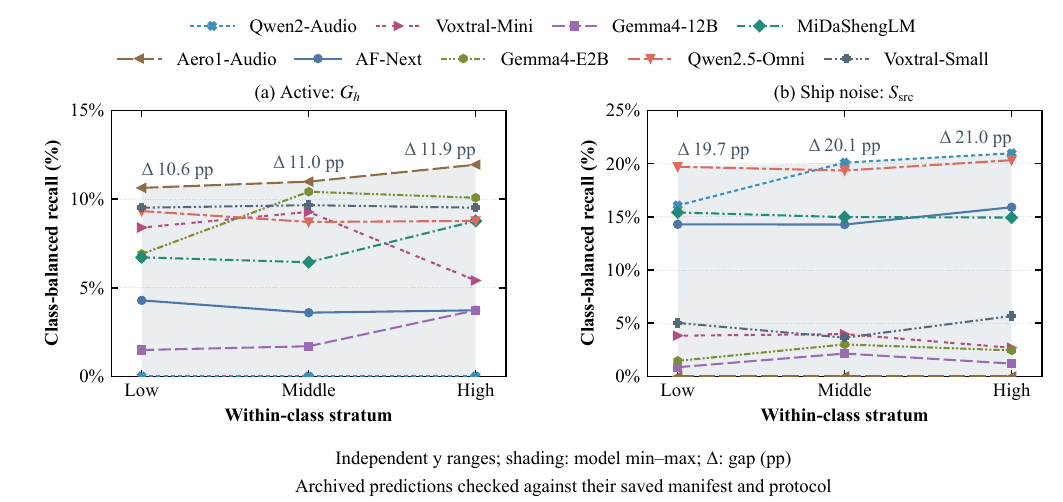}
\caption{\textbf{Metadata-conditioned class-balanced recall under the strict single-letter decoder.} Recall is averaged with equal class weight across (a) the eight active leaves stratified by channel energy gain $G_h$ and (b) the five ship-noise leaves stratified by source tonal-to-continuum power ratio $S_{\mathrm{src}}$. The plotted values reproduce Table~\ref{tab:metadata_results}.}
\label{fig:metadata_performance}
\end{figure}

The strata do not show a shared monotonic pattern. From the low to the high Active stratum, the endpoint difference is $+3.2$ percentage points for Gemma-4-E2B and $-3.0$ points for Voxtral-Mini; both models deviate from a monotonic trend in the middle stratum. Qwen2-Audio has zero Active recall, whereas its ship-noise recall rises from 16.4\% in the low stratum to 21.2\% in the high stratum. The remaining models show smaller or non-monotonic endpoint differences. Figure~\ref{fig:metadata_performance} visualizes the same strict-run values and makes the lack of a common trend apparent across checkpoints.

\subsection{Lexical Explanation Analysis}
\label{sec:explanation_results}

Table~\ref{tab:reasoning} reports the lexical categories defined in Section~\ref{sec:diagnostic_analyses}. The denominator for each row is the number of Turn~3 responses executed after a correct Turn~1 answer. The four categories partition these responses and measure lexical consistency with the predicted leaf. The contradiction rate combines Concept Confusion and Term Stacking.

\begin{table}[!htbp]
\centering
\small
\caption{\textbf{Lexical consistency with predicted leaf classes under the strict single-letter decoder.} The four response categories sum to 100\% up to rounding. Contradiction is the sum of Concept Confusion and Term Stacking. Bold values mark the best model result, and underlining marks the second-best result.}
\label{tab:reasoning}
\begin{tabular*}{\textwidth}{@{\extracolsep{\fill}}lr>{\columncolor{metricaccent}[0pt][\tabcolsep]}cccc>{\columncolor{metricaccent}[0pt][0pt]}c@{}}
\toprule
\tworowhead{Model} & \tworowhead{$n$} & \multicolumn{2}{c}{\textbf{Without competing terms}} & \multicolumn{2}{c}{\textbf{With competing terms}} & \textbf{Summary} \\
\cmidrule(lr){3-4}\cmidrule(lr){5-6}\cmidrule(lr){7-7}
& & \textbf{Aligned$\uparrow$} & \textbf{Vague$\downarrow$} & \textbf{\shortstack{Concept\\confusion$\downarrow$}} & \textbf{\shortstack{Term\\stacking$\downarrow$}} & \textbf{\shortstack{Contradiction\\rate$\downarrow$}} \\
\midrule
Gemma-4-12B-IT~\cite{gemma4} & 717 & \bestcell{87.6} & \secondbest{6.0} & 5.2 & 1.3 & 6.4 \\
Gemma-4-E2B-IT~\cite{gemma4} & 1342 & 37.6 & 59.9 & \secondbest{2.2} & \bestcell{0.2} & \bestcell{2.5} \\
Audio-Flamingo-Next~\cite{audioflamingonext} & 2277 & 20.1 & 29.6 & 46.1 & 4.2 & 50.3 \\
MiDashengLM-7B-1021~\cite{midashenglm} & 1495 & 29.7 & 66.1 & \bestcell{1.6} & 2.6 & \secondbest{4.2} \\
Voxtral-Small-24B-2507~\cite{voxtral_small} & 1476 & 61.0 & \bestcell{5.6} & 22.8 & 10.5 & 33.3 \\
Voxtral-Mini-3B-2507~\cite{voxtral_mini} & 1536 & \secondbest{62.6} & 17.8 & 17.8 & 1.8 & 19.6 \\
Aero-1-Audio~\cite{aero1audio} & 1594 & 6.8 & 71.3 & 21.5 & \secondbest{0.4} & 21.9 \\
Qwen2.5-Omni-7B~\cite{qwen25omni} & 1886 & 58.4 & 25.3 & 6.7 & 9.6 & 16.3 \\
Qwen2-Audio-7B-Instruct~\cite{chu2024qwen2} & 1000 & 37.8 & 8.9 & 19.2 & 34.1 & 53.3 \\
\bottomrule
\end{tabular*}
\end{table}
\FloatBarrier

Gemma-4-12B has the highest aligned rate at 87.6\%, followed by Voxtral-Mini at 62.6\%. This rate is conditional on an executed Turn~3 response: Gemma-4-12B contributes only 717 eligible responses because of its large Turn~1 failure count, so the high conditional rate does not imply reliable end-to-end performance. Voxtral-Small has the lowest vague rate at 5.6\%, while MiDashengLM has the lowest concept-confusion rate at 1.6\%. Lexical alignment therefore varies independently of terminal accuracy and captures a separate property of model responses.

\subsection{Silent-Audio Control}
\label{sec:silent_control_results}

Table~\ref{tab:silent_control} reports the paired original-versus-silent analyses for all nine checkpoints. Each silent waveform has the same duration as its corresponding original input, while the questions and labels remain unchanged. The primary quantity is the original-minus-silent difference: a positive value indicates higher performance with the original waveform than with its paired silent input. The table reports source-cluster bootstrap 95\% confidence intervals; Wilson accuracy intervals and the analytic paired-difference interval are retained in the machine-readable report as cross-checks.

\begin{figure}[!t]
\centering
\includegraphics[
  width=\textwidth,
]{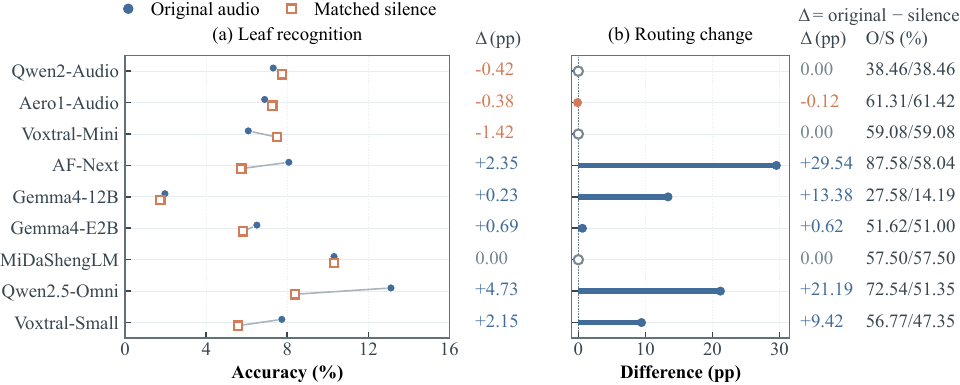}
\caption{\textbf{Paired original-versus-silent control at the leaf and routing levels on UA-Bench-Eval.} Each checkpoint is evaluated on 2,600 original--silent input pairs. (a) L3 leaf-recognition accuracy with original audio (filled blue circles) and duration-matched silence (open orange squares); lines connect the two accuracies for each checkpoint. (b) Original-minus-silent differences in L1 routing accuracy; O/S lists the original/silent accuracies in percent. All $\Delta$ values are computed before rounding and expressed in percentage points (pp). The figure shows point estimates without confidence intervals; L3 accuracy intervals are reported in Table~\ref{tab:silent_control}.}
\label{fig:silent_control_hierarchy}
\end{figure}

\begin{table}[!t]
\centering
\small
\setlength{\tabcolsep}{2.0pt}
\caption{\textbf{Paired original-versus-silent control on UA-Bench-Eval.} All pairs pass the integrity checks. Differences are original minus silent; bracketed ranges are source-cluster bootstrap 95\% confidence intervals based on 10,000 source-level resamples. Accuracy is reported in percent, with differences in percentage points, and Macro F1 on a scale from 0 to 1. The shaded column pairs report the paired effects and their confidence intervals.}
\label{tab:silent_control}
\begin{tabularx}{\textwidth}{@{}>{\raggedright\arraybackslash}X>{\centering\arraybackslash}p{0.070\textwidth}>{\centering\arraybackslash}p{0.060\textwidth}>{\columncolor{metricaccent}[\tabcolsep][\tabcolsep]\centering\arraybackslash}p{0.065\textwidth}>{\columncolor{metricaccent}[\tabcolsep][0pt]\centering\arraybackslash}p{0.140\textwidth}>{\centering\arraybackslash}p{0.070\textwidth}>{\centering\arraybackslash}p{0.060\textwidth}>{\columncolor{metricaccent}[\tabcolsep][\tabcolsep]\centering\arraybackslash}p{0.065\textwidth}>{\columncolor{metricaccent}[\tabcolsep][0pt]\centering\arraybackslash}p{0.140\textwidth}@{}}
\toprule
\tworowhead{Model} & \multicolumn{4}{c}{\textbf{L3 accuracy}} & \multicolumn{4}{c}{\textbf{L3 Macro F1}} \\
\cmidrule(lr){2-5}\cmidrule(lr){6-9}
& \textbf{Original} & \textbf{Silent} & \textbf{$\Delta$} & \textbf{95\% CI} & \textbf{Original} & \textbf{Silent} & \textbf{$\Delta$} & \textbf{95\% CI} \\
\midrule
Gemma-4-12B-IT~\cite{gemma4} & 2.0 & 1.7 & $+0.2$ & \mbox{[$-0.4$, $+0.8$]} & 0.023 & 0.023 & $-0.001$ & \mbox{[$-0.009$, $+0.007$]} \\
Gemma-4-E2B-IT~\cite{gemma4} & 6.5 & 5.8 & $+0.7$ & \mbox{[$+0.04$, $+1.32$]} & 0.049 & 0.046 & $+0.003$ & \mbox{[$-0.004$, $+0.010$]} \\
Audio-Flamingo-Next~\cite{audioflamingonext} & 8.1 & 5.7 & $+2.3$ & \mbox{[$+1.3$, $+3.4$]} & 0.061 & 0.052 & $+0.009$ & \mbox{[$+0.000$, $+0.018$]} \\
MiDashengLM-7B-1021~\cite{midashenglm} & 10.3 & 10.3 & $0.0$ & \mbox{[$0.0$, $0.0$]} & 0.083 & 0.083 & $0.000$ & \mbox{[$0.000$, $0.000$]} \\
Voxtral-Small-24B-2507~\cite{voxtral_small} & 7.7 & 5.6 & $+2.2$ & \mbox{[$+1.2$, $+3.1$]} & 0.056 & 0.035 & $+0.020$ & \mbox{[$+0.012$, $+0.028$]} \\
Voxtral-Mini-3B-2507~\cite{voxtral_mini} & 6.1 & 7.5 & $-1.4$ & \mbox{[$-2.2$, $-0.7$]} & 0.046 & 0.048 & $-0.002$ & \mbox{[$-0.007$, $+0.003$]} \\
Aero-1-Audio~\cite{aero1audio} & 6.9 & 7.3 & $-0.4$ & \mbox{[$-0.9$, $+0.2$]} & 0.033 & 0.028 & $+0.005$ & \mbox{[$+0.001$, $+0.009$]} \\
Qwen2.5-Omni-7B~\cite{qwen25omni} & 13.1 & 8.4 & $+4.7$ & \mbox{[$+3.5$, $+6.0$]} & 0.082 & 0.059 & $+0.023$ & \mbox{[$+0.013$, $+0.033$]} \\
Qwen2-Audio-7B-Instruct~\cite{chu2024qwen2} & 7.3 & 7.7 & $-0.4$ & \mbox{[$-1.1$, $+0.3$]} & 0.063 & 0.064 & $-0.002$ & \mbox{[$-0.008$, $+0.005$]} \\
\bottomrule
\end{tabularx}
\end{table}

Qwen2.5-Omni shows the largest original-minus-silent L3 accuracy difference at 4.7 percentage points, followed by Audio-Flamingo-Next at 2.3 points and Voxtral-Small at 2.2 points. Gemma-4-E2B has a smaller difference of 0.7 points. The cluster-bootstrap intervals for all four positive L3 accuracy differences exclude zero; the Gemma-4-E2B interval is shown to two decimal places so that its positive lower bound remains visible. Voxtral-Mini instead has a $-1.4$-point difference with an interval of $-2.2$ to $-0.7$ points, indicating higher L3 accuracy on silence in this diagnostic comparison. Gemma-4-12B, Aero-1-Audio, and Qwen2-Audio have small L3 accuracy differences whose intervals include zero, while MiDashengLM's L3 accuracy and Macro F1 are unchanged.

Figure~\ref{fig:silent_control_hierarchy} extends this comparison across the label hierarchy, juxtaposing end-to-end L3 leaf-recognition accuracy with coarse L1 routing accuracy for the same nine checkpoints.

The point estimates show that larger gains in L1 routing can coexist with much smaller gains in end-to-end L3 leaf recognition. Audio-Flamingo-Next has an L1 accuracy difference of $+29.54$ percentage points, compared with $+2.35$ points at L3; the corresponding differences for Qwen2.5-Omni are $+21.19$ and $+4.73$ points. Gemma-4-12B provides a further contrast: its L1 difference is $+13.38$ points, whereas its L3 difference is only $+0.23$ points and the L3 confidence interval includes zero. The L1 comparisons are descriptive; no confidence intervals for these differences are shown in the figure.

Together, Table~\ref{tab:silent_control} and Figure~\ref{fig:silent_control_hierarchy} indicate that the observed contribution of the waveform depends on both the model and the level of the label hierarchy. These conclusions are specific to the evaluated configurations and should not be generalized beyond this paired control.

\section{Discussion and Conclusion}
\label{sec:conclusion}

In this work, we connect hierarchical underwater acoustic recognition with controlled source and propagation metadata. Our benchmark combines thirteen leaf classes, source-grouped partitions, and a fixed evaluation suite within one workflow. We use a gated protocol to measure source routing and leaf classification, and we relate model predictions to signal families, decision stages, generation metadata, and explanation vocabulary.

The evaluated models show limited fine-grained recognition on UA-Bench-Eval. Across the nine strict-decoder runs, L3 accuracy ranges from 2.0\% to 13.1\%, and the highest value is 5.4 percentage points above the gated random reference after rounding. MiDashengLM achieves the highest Macro F1 at 0.083, narrowly ahead of Qwen2.5-Omni at 0.082. The model rankings also change across the hierarchy and source families. Voxtral-Small leads on detection pulses, Gemma-4-E2B-IT leads on communication waveforms, and Qwen2.5-Omni leads on ship-radiated noise. No evaluated model provides a consistent advantage across all three families.

Prediction-level results explain part of this variation. Some models fail because their source-routing answer has an invalid format or selects the wrong branch, while others reach the terminal decision but concentrate their predictions on a small number of leaves. Under the strict single-letter decoder, a correct terminal option identifies the full hierarchical path, which makes the stage-wise error partition directly interpretable. Explanation vocabulary supplies another view of response behavior: Gemma-4-12B has the highest conditional lexical alignment rate, but only 717 responses enter this analysis because its Turn~1 failure count is large. The paired silent-audio results further separate model behaviors. Qwen2.5-Omni, Audio-Flamingo-Next, Voxtral-Small, and Gemma-4-E2B show positive L3 differences whose cluster-bootstrap intervals exclude zero, whereas Voxtral-Mini performs better on silence in this diagnostic comparison. The remaining models show small or zero L3 differences. These outcomes provide evidence about model behavior under the present protocol, not a general claim that audio is or is not useful.

The benchmark represents a synthetic distribution defined by its source ranges, propagation settings, evaluation selection, questions, and decoder. Field recordings will be needed to measure transfer beyond this setting. Future work can also expand environmental diversity, add independently validated labels, and compare model decisions under controlled changes to the audio. We provide UA-Bench as a common basis for these studies and for tracking progress in hierarchical underwater acoustic recognition.

\clearpage

\bibliographystyle{unsrtnat}
\bibliography{main}

\clearpage
\appendix

\section{Source Synthesis Settings}
\label{sec:appendix}
\label{sec:appendix_synthesis}

\paragraph{Active signal sampling.} The pulse generator creates 24,000 CW source waveforms and 28,000 source waveforms for each chirp class. The communication generator creates 16,000 source waveforms for each modulation class. These groups contain 160,000 active source waveforms in total. All waveforms use a sampling rate of 16\,kHz. The occupied frequencies remain between 100 and 7900\,Hz.

Table~\ref{tab:active_design_params} summarizes the main sampling rules. Pulse duration depends on the selected subband. Communication duration follows from the symbol count and symbol rate. OFDM also varies the cyclic prefix and the number of subcarriers.

\begin{table}[!htbp]
\centering
\small
\setlength{\tabcolsep}{5pt}
\caption{\textbf{Active-signal synthesis settings by source family.}}
\label{tab:active_design_params}
\begin{tabular}{@{}>{\raggedright\arraybackslash}p{0.13\textwidth}>{\raggedright\arraybackslash}p{0.22\textwidth}>{\raggedright\arraybackslash}p{0.29\textwidth}>{\raggedright\arraybackslash}p{0.29\textwidth}@{}}
\toprule
\textbf{Family} & \textbf{Center-frequency subbands in Hz} & \textbf{Duration or symbol rule} & \textbf{Additional variation} \\
\midrule
\rowcolor{tableaccent}[0pt][0pt]
\multicolumn{4}{@{}l@{}}{\textit{Detection pulses}} \\
CW & \mbox{300--700}, \mbox{700--1500}, \mbox{1500--3500}, \mbox{3500--7500} & 1--10, 0.5--10, 0.05--5, 0.02--2\,s & Single-frequency pulse \\
LFM and HFM & Same four subbands & Same band-dependent duration ranges as CW & Both sweep directions; minimum bandwidth 50, 50, 50, 100\,Hz \\
\rowcolor{tableaccent}[0pt][0pt]
\multicolumn{4}{@{}l@{}}{\textit{Communication waveforms}} \\
2FSK and 4FSK & Same four subbands & 50 to $\max(50,\lfloor30R_s\rfloor)$ symbols at rate $R_s$ & Modulation index 0.5--5 in steps of 0.5 \\
BPSK and QPSK & Same four subbands & Same symbol-count rule as FSK & Raised-cosine roll-off 0 or 0.5--0.9 \\
OFDM & 1500--3500 and 3500--7500 & 1--10 symbols with a 30\,s limit & 512--4096 subcarriers; cyclic-prefix fraction $\{0,0.125,0.25,0.5\}$ \\
\bottomrule
\end{tabular}
\end{table}

\begin{figure}[!t]
\centering
\includegraphics[width=\textwidth]{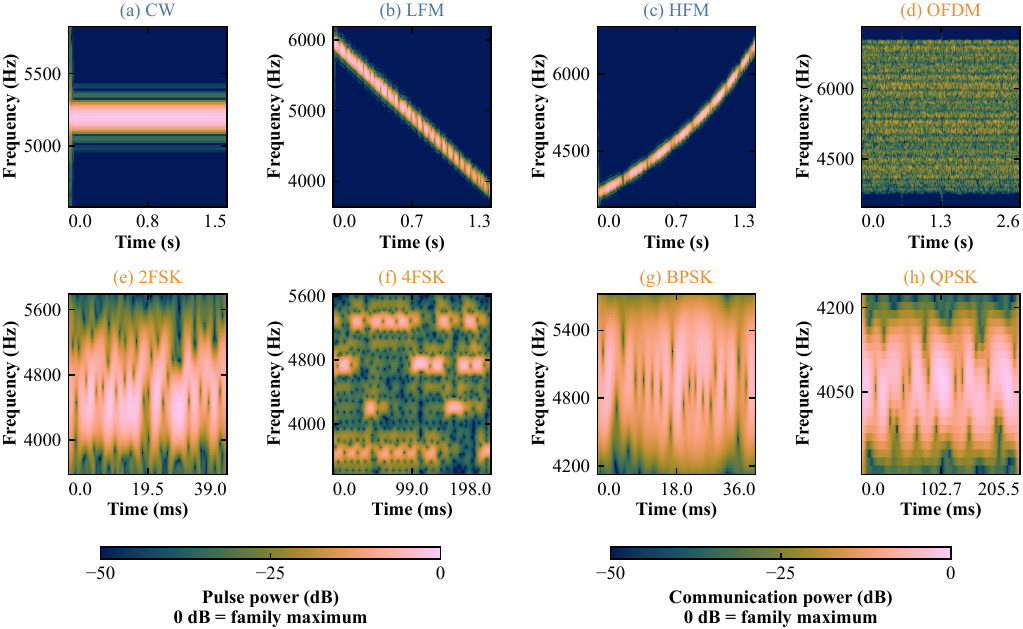}
\caption{\textbf{Representative active-signal structure in UA-Bench-Eval.} (a--c) Time-frequency power for CW, LFM, and HFM detection pulses. (d--h) Local views of OFDM, 2FSK, 4FSK, BPSK, and QPSK communication waveforms. Each panel shows one representative evaluation input with a frequency range and time window matched to the displayed structure. Power is referenced to the maximum within each family, as indicated by the separate color scales.}
\label{fig:acoustic_active}
\end{figure}

Representative evaluation inputs preserve the waveform structures described above. Figure~\ref{fig:acoustic_active} provides the complete active-class gallery used to support the qualitative discussion in Section~\ref{sec:selection}.

\paragraph{Ship-radiated noise sampling.} Each of the five vessel classes contains 16,000 source waveforms. A source waveform combines a shaped random continuum with discrete mechanical tones. Let $b(t)$ denote the continuum. Let $f_{\mathrm{shaft}}$ and $B$ denote the shaft frequency and blade count. The blade-passing frequency is $f_{\mathrm{blade}}=Bf_{\mathrm{shaft}}$. The waveform is
\begin{align}
x(t)&=e(t)b(t)+\sum_{k=1}^{K}a_k\cos(2\pi f_k t+\varphi_k),\\
\widetilde e(t)&=1+m_s\cos(2\pi f_{\mathrm{shaft}}t+\psi_s)
+m_b\cos(2\pi f_{\mathrm{blade}}t+\psi_b)
+\sum_{r=1}^{R}m_r\cos(2\pi f_r t+\psi_r),\\
e(t)&=\frac{\max(\widetilde e(t),0.05)}{\operatorname{mean}_t\max(\widetilde e(t),0.05)}.
\end{align}

The continuum is flat below 300\,Hz and decreases by 6\,dB per octave above that frequency. A 32nd-order autoregressive filter implements this shape. The sampled 1\,kHz spectrum level sets its amplitude. Shaft and blade components share the principal modulation coefficient. One to four additional components use smaller coefficients.

Each source contains between three and ten tones. The generator attempts to associate one to three tones with shaft or blade harmonics. Their frequency jitter is at most $0.3$\,Hz. The remaining tones are sampled within the class range. A spacing of 3\,Hz is enforced when feasible. Each tone has a local line-to-continuum ratio between 6 and 10\,dB in a 1\,Hz reference bandwidth. Table~\ref{tab:ship_full_params} gives the class-specific ranges.

\begin{table}[!htbp]
\centering
\small
\setlength{\tabcolsep}{3.5pt}
\caption{\textbf{Ship-radiated noise settings before propagation.} $B$ denotes blade count and $f_s$ denotes shaft frequency.}
\label{tab:ship_full_params}
\begin{tabularx}{\textwidth}{@{}>{\raggedright\arraybackslash}X*{5}{>{\centering\arraybackslash}p{0.108\textwidth}}@{}}
\toprule
\parbox[c][2.9em][c]{\linewidth}{\raggedright\textbf{Parameter}} &
\parbox[c][2.9em][c]{\linewidth}{\centering\textbf{Underwater\\vehicle}} &
\parbox[c][2.9em][c]{\linewidth}{\centering\textbf{Fishing\\vessel}} &
\parbox[c][2.9em][c]{\linewidth}{\centering\textbf{Cargo\\vessel}} &
\parbox[c][2.9em][c]{\linewidth}{\centering\textbf{Cruise\\ship}} &
\parbox[c][2.9em][c]{\linewidth}{\centering\textbf{Naval\\vessel}} \\
\midrule
\rowcolor{tableaccent}[0pt][0pt]
\multicolumn{6}{l}{\textit{Mechanical configuration}} \\
Blade count $B$ & $\{5,7\}$ & $\{3,4,5\}$ & $\{3,4,5\}$ & $\{3,4,5\}$ & $\{3,4,5\}$ \\
Shaft frequency in Hz & 1.5--18 & 3--35 & 2--25 & 2--22 & 2--28 \\
Blade frequency in Hz & $Bf_s$ & $Bf_s$ & $Bf_s$ & $Bf_s$ & $Bf_s$ \\
Principal modulation coefficient & 0.06--0.08 & 0.15--0.20 & 0.15--0.20 & 0.15--0.20 & 0.10--0.12 \\
\rowcolor{tableaccent}[0pt][0pt]
\multicolumn{6}{l}{\textit{Spectral and tonal configuration}} \\
1\,kHz spectrum level in dB re 1$\mu$Pa$^2$/Hz & 103--130 & 118--136 & 124--142 & 120--148 & 124--148 \\
Continuum corner frequency in Hz & 300 & 300 & 300 & 300 & 300 \\
Continuum slope in dB per octave & $-6$ & $-6$ & $-6$ & $-6$ & $-6$ \\
Number of tones & 3--10 & 3--10 & 3--10 & 3--10 & 3--10 \\
Tone frequency range in Hz & 10--200 & 10--1000 & 10--1000 & 10--1000 & 10--1000 \\
Local line-to-continuum ratio in dB & 6--10 & 6--10 & 6--10 & 6--10 & 6--10 \\
\rowcolor{tableaccent}[0pt][0pt]
\multicolumn{6}{l}{\textit{Geometry and source count}} \\
Source and receiver depth in m & 20--300 & 5--10 & 5--10 & 5--10 & 5--10 \\
Source to receiver range in km & 0.2--3 & 0.2--3 & 0.2--3 & 0.2--3 & 0.2--3 \\
Source waveforms & 16,000 & 16,000 & 16,000 & 16,000 & 16,000 \\
\bottomrule
\end{tabularx}
\end{table}

Blade count is sampled uniformly from the listed choices. Shaft frequency, spectrum level, and principal modulation coefficient are sampled uniformly from their intervals. The internal source class \texttt{underwater\_target} represents a generic propeller-driven underwater platform and is presented to evaluated models under the formal taxonomy name \emph{Underwater vehicle}.

\begin{figure}[!t]
\centering
\includegraphics[width=\textwidth]{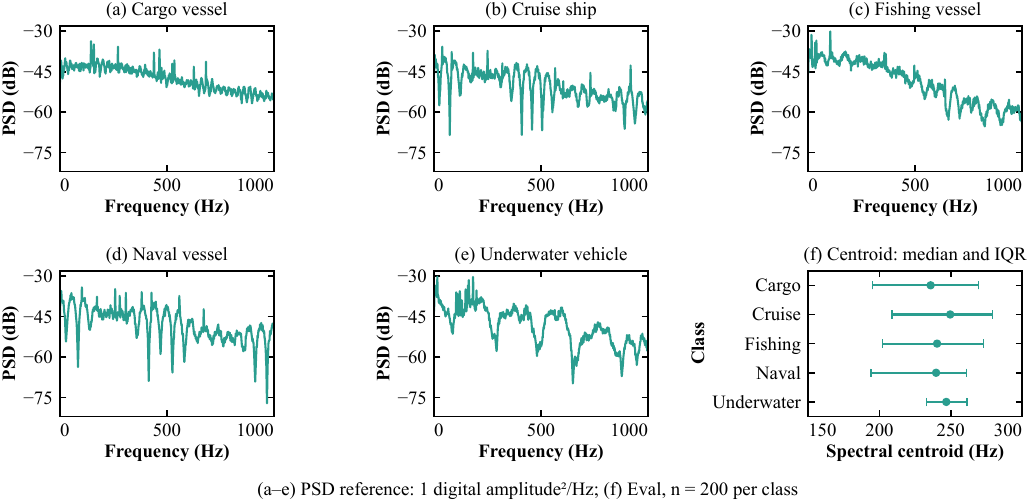}
\caption{\textbf{Representative ship-radiated-noise structure in UA-Bench-Eval.} (a--e) Low-frequency power spectral densities for one evaluation input from each ship-noise leaf class, using a common digital-amplitude-squared-per-hertz reference. (f) Median and interquartile range of spectral centroid over 200 evaluation examples per class. The shortened label \emph{Underwater} denotes the formal taxonomy class \emph{Underwater vehicle}. The spectra illustrate individual inputs rather than class-average templates.}
\label{fig:acoustic_ships}
\end{figure}

Figure~\ref{fig:acoustic_ships} complements the parameter ranges with representative evaluation spectra and the distribution of spectral centroids across the five ship-noise leaf classes.

\FloatBarrier
\section{Propagation Response Construction}
\label{sec:appendix_bellhop}

Each propagation realization selects a sound-speed profile and perturbs the source and receiver geometry. Source depth, receiver depth, and range are perturbed independently by up to 10\%. Source and receiver depths remain within the water column. Table~\ref{tab:bellhop_params} summarizes the shared settings and the differences between source families.

The default number of beams is
\begin{equation}
N_b=\min\{8000,\max\{1000,\lceil100d_{\mathrm{km}}\rceil\}\}.
\end{equation}
If this run returns no arrivals and $N_b<8000$, the simulator repeats it once with $\min\{8000,\max\{4000,2N_b\}\}$ beams. The response metadata records the final beam count and whether the second run was required.

\begin{table}[!htbp]
\centering
\small
\setlength{\tabcolsep}{5pt}
\caption{\textbf{BELLHOP settings for active and ship-radiated sources.} Retained waveforms are scaled to a peak amplitude of 0.95.}
\label{tab:bellhop_params}
\begin{tabularx}{\textwidth}{@{}>{\raggedright\arraybackslash}p{0.23\textwidth}>{\raggedright\arraybackslash}X>{\raggedright\arraybackslash}X@{}}
\toprule
\textbf{Parameter} & \textbf{Active signals} & \textbf{Ship-radiated noise} \\
\midrule
\rowcolor{tableaccent}[0pt][0pt]
\multicolumn{3}{l}{\textit{Response construction}} \\
Response model & Single frequency & Six representative frequencies \\
Frequency in Hz & Signal center or carrier frequency & 50, 100, 200, 400, 800, 1600 \\
Response application & Convolution with one impulse response & Interpolation of complex frequency responses \\
\rowcolor{tableaccent}[0pt][0pt]
\multicolumn{3}{l}{\textit{Environment and numerical settings}} \\
Sound-speed profiles & 100 annual-mean WOA profiles & 100 annual-mean WOA profiles \\
Geometry perturbation & Up to 10\% & Up to 10\% \\
Bottom type & Silt & Silt \\
Bottom sound speed & 1520\,m/s & 1520\,m/s \\
Bottom density & 1.4\,g/cm$^3$ & 1.4\,g/cm$^3$ \\
Bottom attenuation & 3.0\,dB/$\lambda$ & 3.0\,dB/$\lambda$ \\
Beam count & Adaptive from 1000 to 8000 & Adaptive from 1000 to 8000 \\
\rowcolor{tableaccent}[0pt][0pt]
\multicolumn{3}{l}{\textit{Execution and selection}} \\
Propagation attempts per source & 2 & 2 \\
Selection statistic & Impulse-response energy gain & Source tonal-to-continuum power ratio \\
\bottomrule
\end{tabularx}
\end{table}

\paragraph{Arrival response.} An arrival is valid when its delay is finite and nonnegative and its amplitude is finite and positive. The parser retains at most twenty arrivals by amplitude while preserving the earliest arrival. At representative frequency $\nu_q$, the response is
\begin{equation}
h_q[n]=\sum_{p\in\mathcal P_q} A_{qp}\,
k\!\left(n-f_s(\tau_{qp}-\tau_0)\right),
\end{equation}
where $A_{qp}$ and $\tau_{qp}$ are the arrival magnitude and delay. The function $k$ is a 33-tap windowed-sinc kernel with a Kaiser parameter of 6.0. The earliest retained arrival defines $\tau_0$ and provides a common reference for all usable frequency responses in one ship realization.

\paragraph{Applying the response.} Active waveforms are convolved with the response at their center or carrier frequency. For ship-radiated noise, let $H_q(f)$ denote the Fourier transform of $h_q$. The response between adjacent representative frequencies is
\begin{equation}
H(f)=(1-w)H_q(f)+wH_{q+1}(f),\qquad
w=\frac{f-\nu_q}{\nu_{q+1}-\nu_q}.
\end{equation}
The nearest endpoint response is used outside the available frequency range. If only one frequency has usable arrivals, that response is used for the complete waveform. The inverse transform produces a linear convolution, which is cropped to the source frame count. The retained waveform is then scaled to a peak amplitude of 0.95.

\paragraph{Retaining a propagation realization.} An active realization requires one usable response. A ship realization requires at least one usable representative frequency. Both branches reject a nonfinite or zero output waveform. Audio and metadata are written only after these checks succeed.

\section{Metadata, Records, and Dataset Selection}
\label{sec:metadata_provenance}

The active field \texttt{tl\_db} stores the impulse-response energy gain $G_h$ defined in Equation~\ref{eq:quality}. It is computed before output scaling and is not conventional positive transmission loss. The ship field \texttt{snr\_db} stores the source statistic \texttt{snr\_after\_mix\_db}. The metadata converter reconstructs the complete tonal component from its saved frequencies, amplitudes, and phases. It subtracts this component from the source waveform and compares the mean-square powers. This statistic differs from the local ratio used to set each individual tone.

The annotation stage copies the source and channel metadata into each audio record. It also adds the canonical label path and source identity. The channel suffix is removed from the record identifier to recover the source identity. The released fields and selection settings are summarized in Table~\ref{tab:record_fields}.

\begin{table}[!htbp]
\centering
\small
\caption{\textbf{Released record fields and dataset selection settings.} The two panels connect the stored record structure with the source-grouped split and fixed evaluation subset.}
\label{tab:record_fields}
\begin{minipage}[t]{0.58\textwidth}
\centering
\textit{(a) Released record fields}\par\smallskip
\begin{tabularx}{\linewidth}{@{}>{\raggedright\arraybackslash}p{0.34\linewidth}X@{}}
\toprule
\textbf{Field} & \textbf{Content} \\
\midrule
\texttt{id}, \texttt{audio} & Propagation realization identifier and processed waveform path \\
\texttt{conversations} & Three stored questions interleaved with three reference answers \\
\texttt{\_gt} & Canonical L1, L2, and L3 labels \\
\texttt{\_meta} & Source identity, source parameters, channel parameters, and derived statistics \\
\texttt{qa\_prompt\_version} & Version of the stored question specification \\
\texttt{record\_schema\_version} & Version of the record structure \\
\bottomrule
\end{tabularx}
\end{minipage}
\hfill
\begin{minipage}[t]{0.38\textwidth}
\centering
\textit{(b) Partition and evaluation selection}\par\smallskip
{\renewcommand{\arraystretch}{1.37}%
\begin{tabularx}{\linewidth}{@{}>{\raggedright\arraybackslash}p{0.43\linewidth}X@{}}
\toprule
\textbf{Setting} & \textbf{Value} \\
\midrule
Split unit & Source identity \\
Nominal ratio & 70 / 15 / 15 \\
Split seed & 42 \\
Candidate pool & 400 per leaf \\
Evaluation sample & 200 per leaf \\
Leaf classes & 13 \\
Selection seed & 42 \\
\bottomrule
\end{tabularx}
}
\end{minipage}
\end{table}

Source groups are formed within each leaf class. They are shuffled with seed 42 and assigned to training, validation, and test partitions in a nominal ratio of 70, 15, and 15 percent. Integer group counts are obtained with the largest-remainder rule. All propagation realizations of one source inherit the same partition.

UA-Bench-Eval is drawn from the test partition. Records are ranked within each leaf class by the corresponding finite selection statistic. The first 400 records form the candidate pool. The procedure samples 200 records without replacement from each pool with seed 42. It then verifies record identifiers, label paths, metadata, waveform paths, and class counts.

\section{Question and Reference Construction}
\label{sec:appendix_qa}

The question pool contains eight Turn~1 templates, eight Active Turn~2 templates, and six Passive Turn~2 templates. Annotation selects the wording once and stores it in the record. Evaluation reuses the stored questions. All classification templates end with the same instruction.

\begin{promptbox}
Select the option that you consider the best match for the audio. You must choose exactly one of the listed options. Reply with the uppercase option letter only, without explanation or any additional text.
\end{promptbox}

The eight Turn~1 templates are summarized below using lightly simplified renderings. The released records preserve the exact stored wording. Each template is followed by the same two source-category options and the common answer instruction.

\begin{promptbox}
\textbf{Turn~1 question stems}
\begin{enumerate}
\item Determine whether this underwater sound is a deliberately transmitted signal or source-radiated noise.
\item Is the sound intentionally transmitted for detection or communication, or radiated as noise by an operating platform?
\item Classify the sound by its source, either deliberate transmission or platform-radiated noise.
\item Does this audio contain an intentionally transmitted waveform or noise radiated by a vessel or underwater vehicle?
\item Identify the source category of this sound, either a deliberate acoustic transmission or incidental platform noise.
\item Is this sound produced by deliberate signal transmission or by the operation of a vessel or underwater vehicle?
\item Distinguish a deliberately transmitted acoustic signal from noise radiated by an operating platform.
\item Which source category best describes this underwater sound, intentional transmission or source-radiated noise?
\end{enumerate}
\end{promptbox}

\begin{promptbox}
A. Active, deliberately transmitted signal\\
B. Passive, source-radiated noise
\end{promptbox}

The Active branch uses eight second-turn templates, summarized below using lightly simplified renderings. The field $\{$L1$\}$ is replaced by the confirmed source category before the question is stored.

\begin{promptbox}
\textbf{Turn~2 Active question stems}
\begin{enumerate}
\item The signal is $\{$L1$\}$. Determine whether it is a detection pulse or communication signal, and specify the exact type.
\item The signal is $\{$L1$\}$. Identify its subcategory.
\item The signal is $\{$L1$\}$. What is its specific signal type?
\item The source category is $\{$L1$\}$. Determine whether it is a detection pulse or communication signal, and identify the specific type.
\item The signal is confirmed as $\{$L1$\}$. Identify the specific signal type.
\item Confirmed as $\{$L1$\}$. Which specific signal type is it?
\item The first classification step is complete. Further classify this $\{$L1$\}$ signal.
\item Within the $\{$L1$\}$ category, which specific signal type is it?
\end{enumerate}
\end{promptbox}

The Passive branch uses six second-turn templates, again shown as lightly simplified renderings.

\begin{promptbox}
\textbf{Turn~2 Passive question stems}
\begin{enumerate}
\item The signal is $\{$L1$\}$. Identify which type of vessel or underwater vehicle produced this radiated noise.
\item Confirmed as $\{$L1$\}$. Identify the specific type of noise source.
\item The signal is $\{$L1$\}$. What kind of platform emitted this noise?
\item For this $\{$L1$\}$ signal, determine the target type.
\item What type of target is this radiated noise most likely from?
\item Further classify the noise source of this $\{$L1$\}$ signal.
\end{enumerate}
\end{promptbox}

Turn~1 presents the two source categories. Turn~2 presents complete paths within the branch reached at Turn~1. Table~\ref{tab:turn2_options} gives the fixed option mapping. The displayed \emph{Underwater vehicle} option maps to \texttt{underwater\_target}.

\begin{table}[!htbp]
\centering
\small
\caption{\textbf{Fixed classification options.} Turn~1 uses the first two rows. Turn~2 uses the branch-specific rows.}
\label{tab:turn2_options}
\begin{tabular*}{\textwidth}{@{\extracolsep{\fill}}clll@{}}
\toprule
\textbf{Turn} & \textbf{Option} & \textbf{Label level} & \textbf{Displayed choice} \\
\midrule
1 & A & L1 & Active, deliberately transmitted signal \\
1 & B & L1 & Passive, source-radiated noise \\
\midrule
2 Active & A & L2 and L3 & Detection pulse, CW \\
2 Active & B & L2 and L3 & Detection pulse, LFM \\
2 Active & C & L2 and L3 & Detection pulse, HFM \\
2 Active & D & L2 and L3 & Communication signal, 2FSK \\
2 Active & E & L2 and L3 & Communication signal, 4FSK \\
2 Active & F & L2 and L3 & Communication signal, BPSK \\
2 Active & G & L2 and L3 & Communication signal, QPSK \\
2 Active & H & L2 and L3 & Communication signal, OFDM \\
\midrule
2 Passive & A & L2 and L3 & Ship-radiated noise, Cargo vessel \\
2 Passive & B & L2 and L3 & Ship-radiated noise, Cruise ship \\
2 Passive & C & L2 and L3 & Ship-radiated noise, Fishing vessel \\
2 Passive & D & L2 and L3 & Ship-radiated noise, Naval vessel \\
2 Passive & E & L2 and L3 & Ship-radiated noise, Underwater vehicle \\
\bottomrule
\end{tabular*}
\end{table}

The decoder removes surrounding whitespace and accepts only one ASCII letter corresponding to a displayed option; lowercase and uppercase forms are equivalent. It does not recover labels from expanded or free-form answers. An invalid-format or incorrect Turn~1 answer stops the dialogue. After a correct Turn~1 route, each valid Turn~2 letter resolves a unique L2--L3 path. The raw answer and parsing status are retained.

\paragraph{Turn~3 prompt.}

\begin{promptbox}
Briefly explain your previous classification using acoustic cues you can identify in the audio. Focus on the observed cues and how they support your choice. Do not list, compare, or rule out alternative classes, and do not merely repeat the class name or its definition. If the acoustic evidence is insufficient, state this explicitly rather than inventing supporting details.
\end{promptbox}

The first two reference answers contain the correct option letters. The Turn~3 reference combines a source descriptor, a signal-family descriptor, and three acoustic descriptors. The construction procedure selects one of the following sentence patterns.

\begin{promptbox}
\begin{itemize}
\item The signal is $\{$L1$\}$ and exhibits $\{$L2$\}$; observable evidence includes $\{$L3$\}$.
\item Based on its time-frequency and modulation structure, the signal is $\{$L1$\}$ with $\{$L2$\}$; supporting cues are $\{$L3$\}$.
\item Spectral and temporal analysis indicates $\{$L1$\}$ with $\{$L2$\}$, supported by $\{$L3$\}$.
\item Based on its spectral and modulation structure, the signal is $\{$L1$\}$ with $\{$L2$\}$; supporting cues are $\{$L3$\}$.
\end{itemize}
\end{promptbox}

CW, FSK, PSK, and OFDM use descriptors defined by their waveform structures. LFM and HFM use two trajectory descriptors and one direction descriptor selected from the stored sweep direction. Ship-radiated noise uses tonal extent, tone count, harmonic tags, and modulation coefficients.

For vessel sources, a maximum tone frequency at or below 250\,Hz selects the low-frequency description. A maximum at or below 1200\,Hz with a minimum below 250\,Hz selects the low-to-mid-frequency description. Other ranges select the broad-frequency description. At most four tones indicate sparse structure. Five to seven indicate moderate density, and more than seven indicate dense structure. The largest principal modulation coefficient determines weak, moderate, or strong wording at thresholds of 0.09 and 0.14.

Reference answers and metadata are not sent to the model. Evaluation supplies the waveform, stored questions, and the model's earlier answers. The generated Turn~3 response is used for the lexical analysis below.

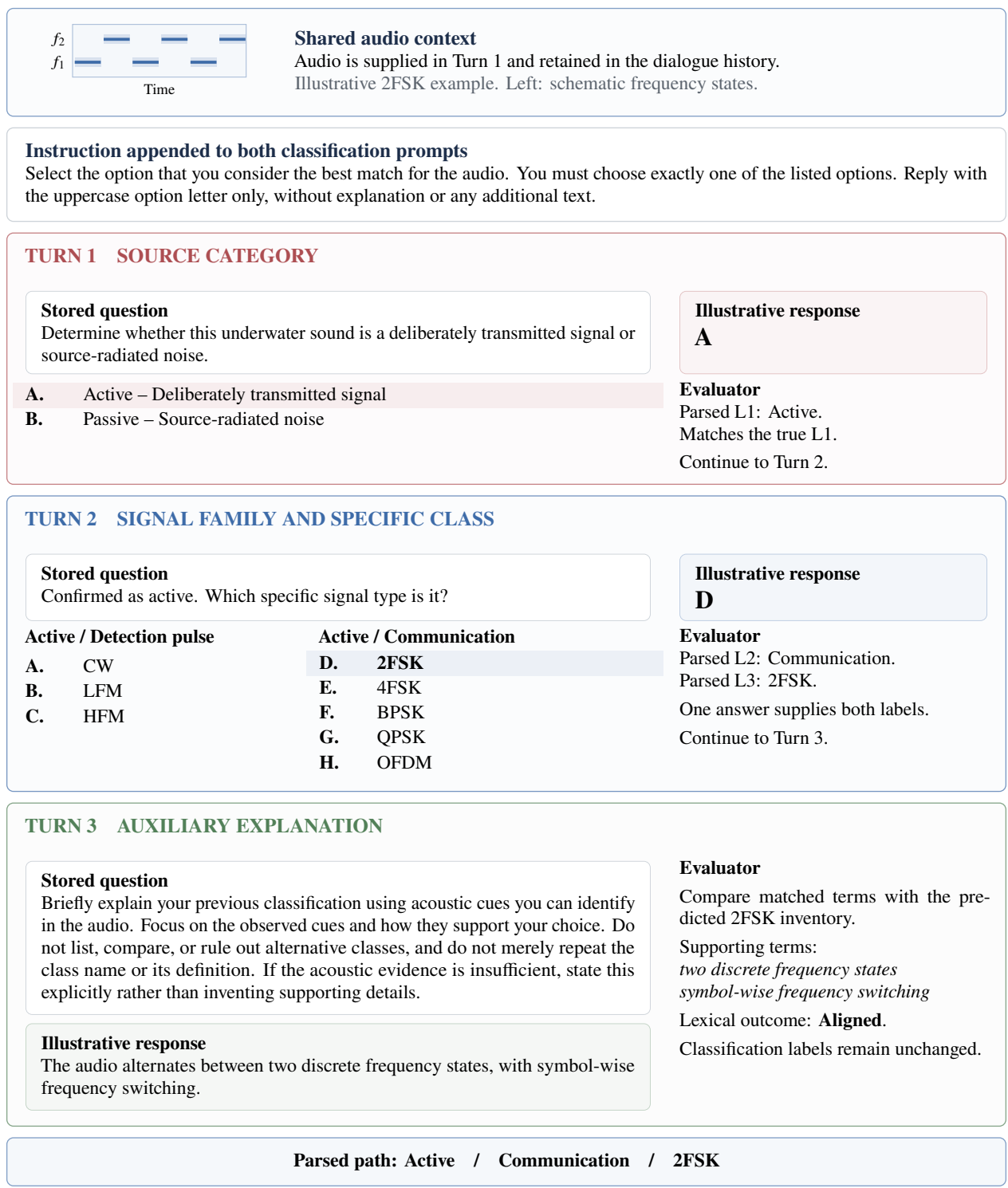
\begin{figure}[!t]
\centering
\begingroup
\definecolor{uaNavy}{RGB}{28,45,73}
\definecolor{uaGray}{RGB}{92,100,110}
\definecolor{uaBorder}{RGB}{205,211,217}
\definecolor{uaCoral}{RGB}{173,78,82}
\definecolor{uaBlue}{RGB}{61,107,168}
\definecolor{uaGreen}{RGB}{82,128,88}
\tcbset{uaCard/.style={enhanced,boxrule=0.5pt,arc=1mm,
  left=2mm,right=2mm,top=1.5mm,bottom=1.5mm,
  before skip=0pt,after skip=5pt,fontupper=\fontsize{9}{10.5}\selectfont},
  uaInner/.style={uaCard,colback=white,colframe=uaBorder,
  boxrule=0.35pt,left=1.5mm,right=1.5mm,top=1mm,bottom=1mm,
  before skip=0pt,after skip=4pt}}

\begin{tcolorbox}[uaCard,colback=uaBlue!2,colframe=uaBlue!65]
\begin{minipage}[c]{0.25\linewidth}
\centering
\begin{tikzpicture}[x=0.24cm,y=0.19cm]
  \fill[uaBlue!5] (0,0) rectangle (12,4.6);
  \foreach \x in {0,4,8}{
    \fill[uaBlue!20] (\x+0.15,0.9) rectangle (\x+1.95,1.7);
    \draw[uaBlue,line width=1.4pt] (\x+0.15,1.3) -- (\x+1.95,1.3);
  }
  \foreach \x in {2,6,10}{
    \fill[uaBlue!20] (\x+0.15,2.9) rectangle (\x+1.95,3.7);
    \draw[uaBlue,line width=1.4pt] (\x+0.15,3.3) -- (\x+1.95,3.3);
  }
  \draw[uaBlue!45] (0,0) rectangle (12,4.6);
  \node[font=\fontsize{7}{8}\selectfont,anchor=east] at (0,1.3) {$f_1$};
  \node[font=\fontsize{7}{8}\selectfont,anchor=east] at (0,3.3) {$f_2$};
  \node[font=\fontsize{7}{8}\selectfont,anchor=north] at (6,0) {Time};
\end{tikzpicture}
\end{minipage}\hfill
\begin{minipage}[c]{0.72\linewidth}
{\fontsize{9.5}{11}\selectfont\bfseries\color{uaNavy}Shared audio context}\par
Audio is supplied in Turn 1 and retained in the dialogue history.\par
{\color{uaGray}Illustrative 2FSK example. Left: schematic frequency states.}
\end{minipage}
\end{tcolorbox}

\begin{tcolorbox}[uaCard,colback=white,colframe=uaBorder]
{\fontsize{9.5}{11}\selectfont\bfseries\color{uaNavy}Instruction appended to both classification prompts}\par
Select the option that you consider the best match for the audio. You must choose exactly one of the listed options. Reply with the uppercase option letter only, without explanation or any additional text.
\end{tcolorbox}

\begin{tcolorbox}[uaCard,colback=uaCoral!2,colframe=uaCoral!70]
{\fontsize{9.5}{11}\selectfont\bfseries\color{uaCoral}TURN 1\quad SOURCE CATEGORY}\par\smallskip
\begin{minipage}[t]{0.65\linewidth}\vspace{0pt}
\begin{tcolorbox}[uaInner,equal height group=turnonehead]
\textbf{Stored question}\par
Determine whether this underwater sound is a deliberately transmitted signal or source-radiated noise.
\end{tcolorbox}
{\renewcommand{\arraystretch}{1.12}%
\begin{tabularx}{\linewidth}{@{}>{\bfseries}p{1.7em}X@{}}
\rowcolor{uaCoral!10} A. & Active -- Deliberately transmitted signal \\
B. & Passive -- Source-radiated noise
\end{tabularx}}
\end{minipage}\hfill
\begin{minipage}[t]{0.32\linewidth}\vspace{0pt}
\begin{tcolorbox}[uaInner,colback=uaCoral!6,colframe=uaCoral!35,equal height group=turnonehead]
\textbf{Illustrative response}\par\smallskip
{\fontsize{12}{13}\selectfont\bfseries A}
\end{tcolorbox}
\textbf{Evaluator}\par
Parsed L1: Active.\par
Matches the true L1.\par\smallskip
Continue to Turn 2.
\end{minipage}
\end{tcolorbox}

\begin{tcolorbox}[uaCard,colback=uaBlue!2,colframe=uaBlue!70]
{\fontsize{9.5}{11}\selectfont\bfseries\color{uaBlue}TURN 2\quad SIGNAL FAMILY AND SPECIFIC CLASS}\par\smallskip
\begin{minipage}[t]{0.65\linewidth}\vspace{0pt}
\begin{tcolorbox}[uaInner,equal height group=turntwohead]
\textbf{Stored question}\par
Confirmed as active. Which specific signal type is it?
\end{tcolorbox}
\begin{minipage}[t]{0.43\linewidth}\vspace{0pt}
\textbf{Active / Detection pulse}\par\smallskip
{\renewcommand{\arraystretch}{1.12}%
\begin{tabularx}{\linewidth}{@{}>{\bfseries}p{1.7em}X@{}}
A. & CW \\
B. & LFM \\
C. & HFM
\end{tabularx}}
\end{minipage}\hfill
\begin{minipage}[t]{0.53\linewidth}\vspace{0pt}
\textbf{Active / Communication}\par\smallskip
{\renewcommand{\arraystretch}{1.12}%
\begin{tabularx}{\linewidth}{@{}>{\bfseries}p{1.7em}X@{}}
\rowcolor{uaBlue!10} D. & \textbf{2FSK} \\
E. & 4FSK \\
F. & BPSK \\
G. & QPSK \\
H. & OFDM
\end{tabularx}}
\end{minipage}
\end{minipage}\hfill
\begin{minipage}[t]{0.32\linewidth}\vspace{0pt}
\begin{tcolorbox}[uaInner,colback=uaBlue!6,colframe=uaBlue!35,equal height group=turntwohead]
\textbf{Illustrative response}\par\smallskip
{\fontsize{12}{13}\selectfont\bfseries D}
\end{tcolorbox}
\textbf{Evaluator}\par
Parsed L2: Communication.\par
Parsed L3: 2FSK.\par\smallskip
One answer supplies both labels.\par\smallskip
Continue to Turn 3.
\end{minipage}
\end{tcolorbox}

\begin{tcolorbox}[uaCard,colback=uaGreen!2,colframe=uaGreen!70]
{\fontsize{9.5}{11}\selectfont\bfseries\color{uaGreen}TURN 3\quad AUXILIARY EXPLANATION}\par\smallskip
\begin{minipage}[t]{0.65\linewidth}\vspace{0pt}
\begin{tcolorbox}[uaInner]
\textbf{Stored question}\par
Briefly explain your previous classification using acoustic cues you can identify in the audio. Focus on the observed cues and how they support your choice. Do not list, compare, or rule out alternative classes, and do not merely repeat the class name or its definition. If the acoustic evidence is insufficient, state this explicitly rather than inventing supporting details.
\end{tcolorbox}
\begin{tcolorbox}[uaInner,colback=uaGreen!6,colframe=uaGreen!35,after skip=0pt]
\textbf{Illustrative response}\par
The audio alternates between two discrete frequency states, with symbol-wise frequency switching.
\end{tcolorbox}
\end{minipage}\hfill
\begin{minipage}[t]{0.32\linewidth}\vspace{0pt}
\textbf{Evaluator}\par\smallskip
Compare matched terms with the predicted 2FSK inventory.\par\smallskip
Supporting terms:\par
\emph{two discrete frequency states}\par
\emph{symbol-wise frequency switching}\par\smallskip
Lexical outcome: \textbf{Aligned}.\par\smallskip
Classification labels remain unchanged.
\end{minipage}
\end{tcolorbox}

\begin{tcolorbox}[uaCard,colback=uaBlue!5,colframe=uaBlue!65,after skip=0pt]
\centering\textbf{Parsed path: Active\quad /\quad Communication\quad /\quad 2FSK}
\end{tcolorbox}
\endgroup
\caption{\textbf{Illustrative three-turn dialogue in UA-Bench.} The 2FSK example follows source routing, complete-path classification, and an auxiliary explanation.}
\label{fig:ua_bench_three_turn_dialogue}
\end{figure}

\FloatBarrier

\section{Lexical Explanation Analysis}
\label{sec:appendix_terms}

Let $V$ be the complete term inventory. Let $S_c\subseteq V$ contain the terms compatible with predicted leaf $c$. Shared cues belong to every compatible leaf. Matching is case insensitive and uses letter and digit boundaries. Candidate phrases are ordered by decreasing length. A phrase is retained when its span does not overlap a previously retained phrase. This rule prevents a shorter phrase from being counted inside a longer match.

Table~\ref{tab:term_sets} summarizes the inventory. The five vessel leaves share the vessel terms and add their displayed class name.

\begingroup
\small
\begin{longtable}{@{}p{0.19\textwidth}p{0.75\textwidth}@{}}
\caption{\textbf{Term inventory used for lexical explanation analysis.}}\label{tab:term_sets}\\
\toprule
\textbf{Class or group} & \textbf{Terms} \\
\midrule
\endfirsthead
\toprule
\textbf{Class or group} & \textbf{Terms continued} \\
\midrule
\endhead
\rowcolor{tableaccent}[0pt][0pt]
\multicolumn{2}{@{}l@{}}{\textit{Shared inventory}} \\
All classes & narrowband, narrowband structure \\
\rowcolor{tableaccent}[0pt][0pt]
\multicolumn{2}{@{}l@{}}{\textit{Detection pulses}} \\
CW & constant frequency, continuous wave, fixed frequency, single tone, single-frequency, stable single-frequency tone, steady frequency, unchanging frequency, unmodulated \\
LFM & chirp, constant sweep rate, frequency decreases linearly with time, frequency increases linearly with time, frequency ramp, linear chirp, linear downsweep, linear frequency modulation, linear sweep, linear upsweep, straight time-frequency trajectory \\
HFM & chirp, curved time-frequency trajectory, frequency decreases nonlinearly with time, frequency increases nonlinearly with time, hyperbolic chirp, hyperbolic frequency modulation, hyperbolic sweep, nonlinear chirp, nonlinear downsweep, nonlinear sweep, nonlinear upsweep, time-varying sweep rate \\
\rowcolor{tableaccent}[0pt][0pt]
\multicolumn{2}{@{}l@{}}{\textit{Communication waveforms}} \\
2FSK & binary FSK, binary frequency-shift keying, frequency switching, symbol-wise frequency switching, two discrete frequency states, two frequency states \\
4FSK & four discrete frequency states, four frequency states, frequency switching, multi-level FSK, quaternary FSK, quaternary frequency-shift keying, symbol-wise frequency switching \\
BPSK & approximately 180-degree phase transitions, binary phase, binary phase-shift keying, binary PSK, phase flip, phase inversion, phase reversal, symbol-wise phase-state transitions, two phase states \\
QPSK & approximately 180-degree phase transitions, four phase states, four-phase, phase flip, phase inversion, phase reversal, quadrature modulation, quadrature phase-shift keying, quadrature PSK, symbol-wise phase-state transitions \\
OFDM & dense carriers, multi-carrier, multiple subcarriers, orthogonal carriers, orthogonal frequency-division multiplexing, regularly spaced subcarriers, simultaneous multicarrier structure, subcarriers \\
\rowcolor{tableaccent}[0pt][0pt]
\multicolumn{2}{@{}l@{}}{\textit{Ship-radiated noise}} \\
Shared vessel terms & blade-rate-related harmonic lines, broadband-plus-tonal vessel noise, dense discrete tonal structure, discrete machinery tonal components, low-frequency tonal structure, mechanical tonal components without a dominant harmonic family, moderate periodic amplitude modulation, moderate-density discrete tonal structure, periodic propeller-related modulation, shaft- and blade-rate-related harmonic lines, shaft-rate-related harmonic lines, sparse discrete tonal structure, strong periodic amplitude modulation, tonal components concentrated in the low-frequency band, tonal components spanning a broad frequency range, tonal components spread across the low-to-mid-frequency band, weak periodic amplitude modulation, weak periodic propeller-related modulation \\
Cargo vessel & cargo vessel \\
Cruise ship & cruise ship \\
Fishing vessel & fishing vessel \\
Naval vessel & naval vessel \\
Underwater vehicle & underwater vehicle \\
\bottomrule
\end{longtable}
\endgroup

For response $i$, let $p_i$ indicate at least one retained match in $S_{\hat y_i^{(3)}}$. Let $q_i$ indicate at least one retained match in its complement. If $M$ Turn~3 responses are executed, the rates are
\begin{align}
R_{\mathrm{aligned}}&=M^{-1}\sum_i p_i(1-q_i), &
R_{\mathrm{confusion}}&=M^{-1}\sum_i (1-p_i)q_i,\\
R_{\mathrm{vague}}&=M^{-1}\sum_i(1-p_i)(1-q_i), &
R_{\mathrm{stacking}}&=M^{-1}\sum_i p_iq_i,\\
R_{\mathrm{contradiction}}&=M^{-1}\sum_i q_i
=R_{\mathrm{confusion}}+R_{\mathrm{stacking}}.
\end{align}

The first four categories partition the executed responses. An invalid-format Turn~2 prediction has an empty supporting set and uses $V$ as its comparison set. Such a response is Concept Confusion when an inventory term is present and Vague otherwise. An empty explanation is also Vague.

\section{Classification Metrics and Random References}
\label{sec:appendix_metrics}

For leaf class $c$, precision, recall, and F1 are
\begin{equation}
P_c=\frac{TP_c}{TP_c+FP_c},\qquad
R_c=\frac{TP_c}{TP_c+FN_c},\qquad
F_c=\frac{2P_cR_c}{P_c+R_c}.
\end{equation}
Invalid-format and cascade predictions contribute false negatives to their true classes. They also appear as separate prediction outcomes in the confusion matrix. L3 Macro F1 is the mean F1 over the thirteen leaf classes.

A ratio with an empty denominator is recorded as zero. The evaluator rounds each per-class precision, recall, and F1 value to four decimal places. Macro F1 is the unweighted mean of the thirteen stored four-decimal class F1 values. Runs without an executed Turn~3 response record a zero response count and zero lexical rates.

For a valid Turn~2 option, leaf correctness implies correct source and signal-family labels. Consequently, L3 accuracy equals complete-path accuracy under the strict decoder and no duplicate joint metric is reported.

The random reference chooses each source category with probability $1/2$. On the Active branch, three of the eight options belong to detection pulses and five belong to communication signals. The Passive branch contains one signal family. The expected L2 accuracy is
\begin{equation}
\mathbb E[\mathrm{Acc}_{L2}]
=\frac3{13}\frac12\frac38
+\frac5{13}\frac12\frac58
+\frac5{13}\frac12
=\frac{37}{104}\simeq35.58\%.
\end{equation}
The corresponding conditional references are $37/52\simeq71.15\%$ for L2$|$L1 and $8/37\simeq21.62\%$ for L3$|$L2. Together with L1 accuracy of $1/2$, they yield the $1/13$ leaf reference in Equation~\ref{eq:random}.

For the paired silent-audio control, the original and silent L3 accuracies receive Wilson score intervals. The original-minus-silent accuracy difference also receives an analytic paired-proportion interval based on the two discordant-pair counts. The reported 95\% intervals for original accuracy, silent accuracy, their difference, and the corresponding Macro F1 quantities use a percentile cluster bootstrap. One cluster is one source identity, obtained by removing a terminal \texttt{\_chN} channel suffix from the sample identifier. Channel variants are therefore resampled together. The evaluator uses 10,000 replicates and seed 20261008 by default.

\FloatBarrier
\section{Inference and Reproducible Evaluation}
\label{sec:appendix_repro}

We identify every evaluated checkpoint by its repository namespace and checkpoint identifier in Table~\ref{tab:model_suite_status}. Across these backends, the shared execution, input, and preprocessing settings are collected in Table~\ref{tab:model_config}. Each backend uses its native processor and chat template. The waveform enters the first user turn and remains in the dialogue history. Later turns use the questions stored in each evaluation record.

\begin{table}[!htbp]
\centering
\small
\caption{\textbf{Evaluated model repositories and checkpoint identifiers.}}
\label{tab:model_suite_status}
\begin{tabularx}{\textwidth}{@{}>{\raggedright\arraybackslash}p{0.31\textwidth}>{\raggedright\arraybackslash}p{0.15\textwidth}>{\raggedright\arraybackslash}X@{}}
\toprule
\textbf{Model} & \textbf{Repository} & \textbf{Checkpoint identifier} \\
\midrule
Gemma-4-12B-IT~\cite{gemma4} & \texttt{google} & \texttt{gemma-4-12B-it} \\
Gemma-4-E2B-IT~\cite{gemma4} & \texttt{google} & \texttt{gemma-4-E2B-it} \\
Audio-Flamingo-Next~\cite{audioflamingonext} & \texttt{nvidia} & \texttt{audio-flamingo-next-hf} \\
MiDashengLM-7B-1021~\cite{midashenglm} & \texttt{mispeech} & \texttt{midashenglm-7b-1021-bf16} \\
Voxtral-Small-24B-2507~\cite{voxtral_small} & \texttt{mistralai} & \texttt{Voxtral-Small-24B-2507} \\
Voxtral-Mini-3B-2507~\cite{voxtral_mini} & \texttt{mistralai} & \texttt{Voxtral-Mini-3B-2507} \\
Aero-1-Audio~\cite{aero1audio} & \texttt{lmms-lab} & \texttt{Aero-1-Audio} \\
Qwen2.5-Omni-7B~\cite{qwen25omni} & \texttt{Qwen} & \texttt{Qwen2.5-Omni-7B} \\
Qwen2-Audio-7B-Instruct~\cite{chu2024qwen2} & \texttt{Qwen} & \texttt{Qwen2-Audio-7B-Instruct} \\
\bottomrule
\end{tabularx}
\end{table}

\begin{table}[!htbp]
\centering
\small
\caption{\textbf{Shared inference settings without task demonstrations.}}
\label{tab:model_config}
\begin{tabularx}{\textwidth}{@{}>{\raggedright\arraybackslash}p{0.18\textwidth}>{\raggedright\arraybackslash}p{0.27\textwidth}>{\raggedright\arraybackslash}p{0.17\textwidth}>{\raggedright\arraybackslash}X@{}}
\toprule
\multicolumn{2}{c}{\textbf{Generation and execution}} & \multicolumn{2}{c}{\textbf{Input and preprocessing}} \\
\cmidrule(lr){1-2}\cmidrule(lr){3-4}
\textbf{Parameter} & \textbf{Value} & \textbf{Parameter} & \textbf{Value} \\
\midrule
Numerical precision & bfloat16 where supported & Benchmark waveform & Mono at 16\,kHz \\
Maximum text output & 256 new tokens per turn & Model preprocessing & Native model processor \\
Text decoding & Greedy decoding where supported & Batch size & Backend-specific \\
Audio output & Disabled & Evaluation questions & Stored questions from each record \\
\bottomrule
\end{tabularx}
\end{table}

Each prediction record preserves the questions, raw answers, parsed labels, and turn status. Each protocol file records the model settings, taxonomy, evaluation manifest hash, and implementation hashes. Aggregation requires complete shard coverage and a shared run identity. It also checks sample coverage and duplicate identifiers before calculating metrics.

Metadata strata use empirical one-third and two-thirds quantiles within each leaf class. A value equal to a boundary enters the higher stratum. Recall is computed for each leaf and stratum. The branch result is the unweighted mean of its leaf recalls. The evaluation records and thresholds remain fixed across models.

\FloatBarrier

\end{document}